\documentclass[sigconf]{acmart}

\acmConference[Arxiv]{arxiv.org}{Feb 2022}{USA}

\usepackage[T1]{fontenc}
\usepackage{paralist}
\usepackage{newunicodechar}
\usepackage{fontawesome}
\usepackage{wrapfig}
\usepackage{tabularx}

\usepackage{balance}
\usepackage{booktabs,dcolumn}
\usepackage{multirow}
\usepackage{placeins}

\newcommand\myparagraph[1]{\vspace{.4em}\noindent\textbf{#1}\ }
\renewcommand\footnotetextcopyrightpermission[1]{}

\newcommand\papertitle{Collaboration Challenges in Building ML-Enabled Systems: Communication, Documentation, Engineering, and Process}
\newcommand\paperabstract{The introduction of machine learning (ML) components in software projects has created the need for software engineers to collaborate with data scientists and other specialists. While collaboration can always be challenging, ML introduces additional challenges with its exploratory model development process, additional skills and knowledge needed, difficulties testing ML systems, need for continuous evolution and monitoring, and non-traditional quality requirements such as fairness and explainability. Through interviews with 45 practitioners from 28 organizations, we identified key collaboration challenges that teams face when building and deploying ML systems into production. We report on common collaboration points  in the development of production ML systems for requirements, data, and integration, as well as corresponding team patterns and challenges. We find that most of these challenges center around communication, documentation, engineering, and process, and collect recommendations to address these challenges.}

\begin{document}
\title{\papertitle}
	\author{Nadia Nahar}
\email{nadian@andrew.cmu.edu}
\affiliation{%
   \institution{Carnegie Mellon University}
   \city{Pittsburgh}
   \state{PA}
   \country{USA}}
\author{Shurui Zhou}
\affiliation{%
   \institution{University of Toronto}
   \city{Toronto}
   \state{Ontario}
   \country{Canada}}
\author{Grace Lewis}
\affiliation{%
   \institution{Carnegie Mellon Software Engineering Institute}
   \city{Pittsburgh}
   \state{PA}
   \country{USA}}
\author{Christian Kästner}
\affiliation{%
   \institution{Carnegie Mellon University}
   \city{Pittsburgh}
   \state{PA}
   \country{USA}}
\begin{abstract}
	\paperabstract
\end{abstract}
		\maketitle

	\label{h.1xopwa74r7xw}

\section{Introduction}

\label{h.7njfo9tpsac}
\begin{figure}[t]
 \vspace{0.1cm}
\includegraphics[width=.88\linewidth]{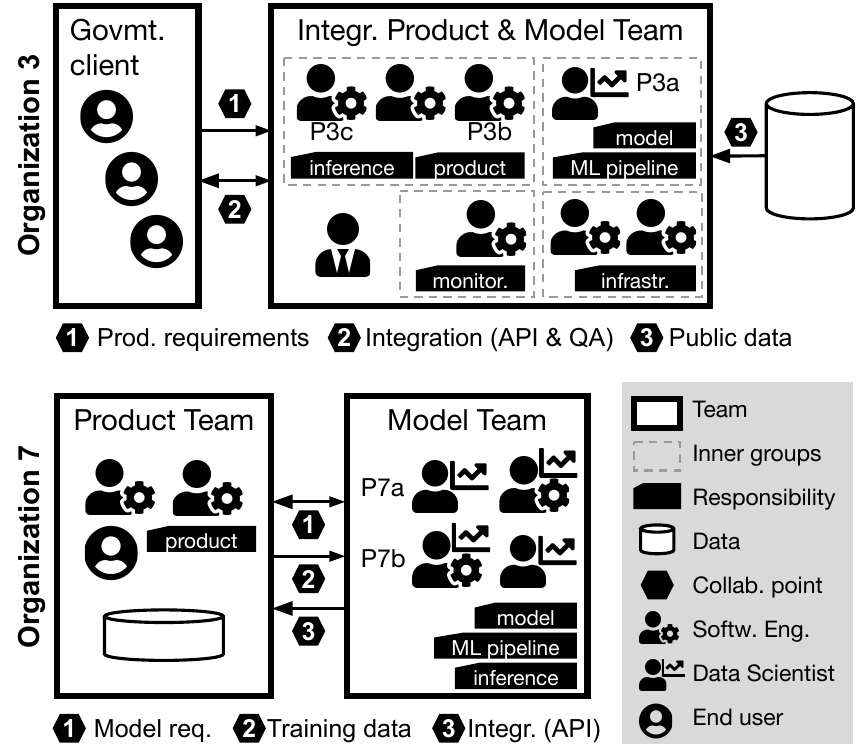}
\vspace{-0.2cm}
\caption{Structure of two interviewed organizations}
\label{fig:orgstructure}
\end{figure}

\newcommand\tC{}

Machine learning (ML) is receiving massive attention and funding in research and practice; it is achieving incredible advances, surpassing human-level cognition in many applications, but it is widely acknowledged that moving from a prototyped machine-learned model to a production system is very challenging. For example, Venturebeat reported in 2019 that 87~percent of ML projects fail \cite{sr9P} and Gartner claimed in 2020 that 53~percent do not make it from prototype to production \cite{GYwY}. While traditional software projects are already complex, failure prone, and require a broad range of expertise, the introduction of machine learning raises further challenges, requires additional expertise, and introduces additional collaboration points.

Technical aspects such as testing ML components  \cite{Pd6Q,i8ow}, misuse of ML libraries \cite{gXfW,efZw}, engineering challenges for developing ML components \cite{FdO8,7sej,gC0G,VVrZ,tPQ5,UjZF,NhgE,6ckN}, and automating learning and deployment processes for ML components  \cite{Kol9,5guS,I17M,6aXp,5pvj}, have received significant attention in research recently. However, human factors of collaboration during the development of software products supported by ML components, \emph{ML-enabled systems} for short, have received less attention, including the need to separate and coordinate data science and software engineering work, to negotiate and document interfaces and responsibilities, and to plan the system's operation and evolution. Yet, those human collaboration challenges appear to be major hurdles in developing ML-enabled systems. In addition, past work has mostly been model-centric, focused on challenges of learning, testing, or serving models, but rarely focuses on the entire system, i.e., the product with many non-ML parts into which the model is embedded as a component, which requires coordinating and integrating work from multiple experts or teams.

To better understand collaboration challenges and avenues toward better practices, we conducted interviews with 45 participants contributing to the development of ML-enabled systems for production use (i.e., not pure data analytics/early prototypes). Our research question is: \emph{What are the collaboration points and corresponding challenges between data scientists and software engineers?} Participants come from 28 organizations, from small startups to large big tech companies, and have diverse roles in these projects, including data scientists, software engineers, and managers. During our interviews, we explored organizational structures (e.g., see Figure~\ref{fig:orgstructure}), interactions of project members with different technical backgrounds, and where conflicts arise between teams. \looseness=-1

While some organizations have adopted better collaboration practices than others, many struggle setting up structures, processes, and tooling for effective collaboration among team members with different backgrounds when developing ML-enabled systems. To the best of our knowledge, and confirmed by the practitioners we interviewed, there is little systematic or shared understanding of common collaboration challenges and best practices for developing ML-enabled systems and coordinating developers with very different backgrounds (e.g., data science vs. software engineering). We find that smaller and new-to-ML organizations struggle more, but have limited advice to draw from for improvement.

Three \emph{collaboration points} surfaced as particularly challenging: (1)~Identifying and decomposing requirements, (2)~negotiating training data quality and quantity, and (3)~integrating data science and software engineering work. We found that organizational structure, team composition, power dynamics, and responsibilities differ substantially, but also found common \emph{organizational patterns} at specific collaboration points and challenges associated with them. 

Overall, our observations suggest four \emph{themes} that would benefit from more attention when building ML-enabled systems: \faGroup~Invest in supporting \emph{interdisciplinary teams} to work together (including education and avoiding silos), \faFileText~Pay more attention to collaboration points and clearly \emph{document} responsibilities and interfaces, \faGears~Consider \emph{engineering work} as a key contribution to the project, and \faCalendar~Invest more into \emph{process and planning}.

In summary, we make the following contributions: (1) We identify three core collaboration points and associated collaboration challenges based on interviews with 45 practitioners, triangulated with a literature review, (2) We highlight the different ways in which teams organize, but also identify organizational patterns that associate with certain collaboration challenges, and (3) We identify recommendations to improve collaboration practices.

\section{State of the Art}
\label{h.kpwvxerzqjbu}

Researchers and practitioners have discussed whether and how machine learning changes software engineering with the introduction of learned models as components in software systems~\cite[e.g.,][]{hpbm,tPQ5,ctC1,aFTn,p7yN,UjZF,vhGB,5Nhu,aho}. To lay the foundation for our interview study and inform the questions we ask, we first provide an overview of the related work and existing theories on collaboration in traditional software engineering and discuss how machine learning may change this.

\myparagraph{Collaboration in Software Engineering.}
\label{h.a7p9iwfygbws}
Most software projects exceed the capacity of a single developer, requiring multiple developers and teams to collaborate (``work together'') and coordinate (``align goals''). Collaboration happens \emph{across} teams, often in a more formal and structured form, and \emph{within} teams, where familiarity with other team members and frequent co-location fosters informal communication \cite{hATC}.  At a technical level, to allow multiple developers to work together, \emph{abstraction} and a \emph{divide and conquer} strategy are essential. Dividing software into \emph{components} (modules, functions, subsystems) and hiding internals behind \emph{interfaces} is a key principle of modular software development that allows teams to divide work, and work mostly independently until the final system is integrated \cite{VII1,06Iv}.

Teams within an organization tend to align with the technical structure of the system, with individuals or teams assigned to components \cite{mFZV}, hence the technical structure (interfaces and dependencies between components) influences the points where teams collaborate and coordinate. Coordination challenges are especially observed when teams cannot easily and informally communicate, often studied in the context of distributed teams of  global corporations \cite{AzHU,FlkT} and open-source ecosystems \cite{1i9g,ozxl}.

More broadly, \emph{interdisciplinary} collaboration often poses challenges. It has been shown that when team members differ in their academic and professional backgrounds and possess different expectations on the same system, communication, cultural, and methodical challenges often emerge when working together \cite{mnPb,passi}. Key insights are that successful interdisciplinary collaboration depends on professional role, structural characteristics, personal characteristics, and a history of collaboration; specifically, structural factors such as unclear mission, insufficient time, excessive workload, and lack of administrative support are barriers to collaboration \cite{N1oU}.

The component \emph{interface} plays a key role in collaboration as a \emph{negotiation and collaboration point}. It is where teams (re-)negotiate how to divide work and assign responsibilities \cite{zu7r}. Team members often seek information that may not be captured in interface descriptions, as interfaces are rarely fully specified \cite{JztF}. In an idealized development process, interfaces are defined early based on what is assumed to remain stable \cite{VII1}, because changes to interfaces later are expensive and require the involvement of multiple teams. In addition, interfaces reflect \emph{key architectural decisions} for the system, aimed to achieve desired overall qualities \cite{FvQJ}.

In practice though, the idealized divide-and-conquer approach following top-down planning does not always work without friction. Not all changes can be anticipated, leading to later modifications and renegotiation of interfaces \cite{4JFg,1i9g}. It may not be possible to identify how to decompose work and design stable interfaces until substantial experimentation has been performed \cite{Ler4}. To manage, negotiate, and communicate changes of interfaces, developers have adopted a wide range of strategies for communication \cite{1i9g,GqTH,MuoQ}, often relying on informal broadcast mechanisms to share planned or performed changes with other teams.

\emph{Software lifecycle models}~\cite{sb6P} also address this tension of when and how to design stable interfaces: Traditional top-down models (e.g., waterfall) plan software design after careful requirements analysis; the \emph{spiral model} pursues a risk-first approach in which developers iterate to prototype risky parts, which then informs future system design iterations; \emph{agile }approaches de-emphasize upfront architectural design for fast iteration on incremental prototypes. The software architecture community has also grappled with the question of how much upfront architectural design is feasible, practical, or desirable \cite{FvQJ,WTXV}, showing a tension between the desire for upfront planning on one side and technical risks and unstable requirements on the other. \emph{In this context, our research explores how introducing machine learning into software projects challenges collaboration.
}

\myparagraph{Software Engineering with ML Components.}
\label{h.s5bm0a4lxbfu}
In a ML-enabled system, machine learning contributes one or multiple components to a larger system with traditional non-ML components. We refer to the whole system that an end user would use as the \emph{product}.  In some systems, the learned \emph{model} may be a relatively small and isolated addition to a large traditional software system (e.g., audit prediction in tax software); in others it may provide the system's essential core with only minimal non-ML code around it (e.g., a sales prediction system sending daily predictions by email). In addition to models, an ML-enabled system typically also  has components for training  and monitoring  the model(s) \cite{hpbm,Kol9}. Much attention in practice recently focuses on building robust ML \emph{pipelines} for training and deploying models in a scalable fashion, often under names such as \emph{``AI engineering,'' ``SysML,''} and \emph{``MLOps''} \cite{Kol9,UjZF,5hJ9,ZfhU}. In this work, we focus more broadly on the development of the entire ML-enabled system, including both ML and non-ML components.

Compared to traditional software systems, ML-enabled systems require additional expertise in \emph{data science} to build the models and may place additional emphasis on expertise such as data management, safety, and ethics \cite{JnkY,tPQ5}. In this paper, we primarily focus on the roles of \emph{software engineers} and \emph{data scientists}, who typically have different skills and educational backgrounds \cite{eIMb,5Nhu,WVIq,JnkY}: Data science education tends to focus more on statistics, ML algorithms, and practical training of models from data (typically given a fixed dataset, not deploying the model, not building a system), whereas software engineering education focuses on engineering tradeoffs with competing qualities, limited information, limited budget, and the construction and deployment of systems. Research shows that software engineers who engage in data science without further education are often naive when building models \cite{5Nhu} and that data scientists prefer to focus narrowly on modeling tasks \cite{WVIq} but are frequently faced with engineering work \cite{bU0N}. While there is plenty of work on supporting collaboration among software engineers \cite{MuoQ,2BW8,03Lt,pevg} and more recently on supporting collaboration among data scientists \cite{U2s5,7LH3}, we are not aware of work exploring collaboration challenges between these roles, which we do in this work. \looseness=-1

The software engineering community has recently started to explore \emph{software engineering for ML-enabled systems} as a research field, with many contributions on bringing software-engineering techniques to ML tasks, such as testing models  and ML algorithms \cite{WaYh,bQ59,Pd6Q,i8ow}, deploying models \cite{Kol9,5guS,I17M,6aXp,5pvj}, robustness and fairness  of models \cite{XRxi,8A28,p7yN}, life cycles  for ML models \cite{5guS,sIUl,tPQ5,B5qD,aho}, and  engineering challenges or best practices for developing ML components \cite{FdO8,7sej,gC0G,VVrZ,tPQ5,UjZF,NhgE,6ckN}. A smaller body of work focuses on the ML-enabled system beyond the model, such as exploring system-level quality attributes \cite{Dzd7,passi}, requirements engineering \cite{aFTn}, architectural design \cite{bdD7}, safety mechanisms \cite{We4r,vhGB}, and user interaction design \cite{uoFv,TIPJ,yang}. \emph{In this paper, we adopt this system-wide scope and explore how data scientists and software engineers work together to build the system with }\emph{ML and non-ML components.}

\section{Research Design}

\label{h.234t9x7j3o11}
Because there is limited research on collaboration in building ML-enabled systems, we adopt a qualitative research strategy to explore \emph{collaboration points} and corresponding \emph{challenges}, primarily with stakeholder interviews. We proceeded in four steps: (1)~We prepared interviews based on an initial literature review, (2)~we conducted interviews, (3)~we triangulated results with literature findings, and (4)~we validated our findings with the interview participants. We base our research design on \emph{Straussian Grounded Theory} \cite{ia1C,Hc0J}, which derives research questions from literature, analyzes interviews with open and axial coding, and consults literature throughout the process. In particular, we conduct interviews and literature analysis in parallel, with immediate and continuous data analysis, performing constant comparisons, and refining our codebook and interview questions throughout the study.

\myparagraph{Step 1: Scoping and interview guide.} To scope our research and prepare for interviews, we looked for collaboration problems mentioned in existing literature on software engineering for ML-enabled systems (Sec. \ref{h.kpwvxerzqjbu}). In this phase, we selected 15 papers opportunistically through keyword search and our own knowledge of the field. We marked all sections in those papers that potentially relate to collaboration challenges between team members with different skills or educational backgrounds, following a standard open coding process \cite{ia1C}. Even though most papers did not talk about problems in terms of collaboration, we marked discussions that may plausibly relate to collaboration, such as data quality issues between teams. We then analyzed and condensed these codes into nine initial collaboration areas and developed an initial codebook and interview guide (provided in Supplement B and C at the end).

\myparagraph{Step 2: Interviews.} We conducted semi-structured interviews with 45 participants from 28 organizations, each 30 to 60 minutes long. All participants are involved in professional software projects using machine learning that are either already or planned to be deployed in production. In Table~\ref{demographics}, we show the demographics of the interview participants and their organizations. Details can be found in the Supplement A at the end.

We tried to sample participants purposefully (maximum variation sampling \cite{c7gw}) to cover participants in different roles, types of companies, and countries. We intentionally recruited most participants from organizations outside of big tech companies, as they represent the vast majority of projects that have recently adopted machine learning and often face substantially different challenges \cite{gC0G}. Where possible, we tried to separately interview multiple participants in different roles within the same organization to get different perspectives. We identified potential participants through personal networks, ML-related networking events, LinkedIn, and recommendations from previous interviewees and local tech leaders. We adapted our recruitment strategy throughout the research based on our findings, at later stages focusing primarily on specific roles and organizations to fill gaps in our understanding, until reaching saturation. For confidentiality, we refer to organizations by number and to participants by P\emph{Xy} where \emph{X} refers to the organization number and \emph{y} distinguishes participants from the same organization.

\begin{table}[t]
    \caption{Participant and Company Demographics}
    \vspace{-.5em}
    \label{demographics}
    \small
\begin{tabularx}{\linewidth}{lX}
\toprule
\textbf{Type} & \textbf{Break-down} \\\midrule
Participant Role (45)             & ML-focused (23), SE-focused (9), Management (5), Operations (2), Domain Expert (2), Other~(4)                              \\

Participant Seniority (45)             & 5 years of experience or more~(28), 2-5 years~(9), under 2 years~(8)                             \\
Company Type (28)            & Big tech (6), Non IT (4), Mid-size tech~(11), Startup (5), Consulting (2)                               \\
Company Location (28)                   & North America (11), South America (1), Europe (5), Asia (10), Africa (1) \\\bottomrule
\end{tabularx}
\end{table}

We transcribed and analyzed all interviews. Then, to map challenges to collaboration points, we created visualizations of organizational structure and responsibilities in each organization (we show two examples in Figure~\ref{fig:orgstructure}) and mapped collaboration problems mentioned in the interviews to collaboration points within these visualizations. We used these visualizations to further organize our data; in particular, we explored whether collaboration problems associate with certain types of organizational structures.

\myparagraph{Step 3: Triangulation with literature.} As we gained insights from interviews, we returned to the literature to identify related discussions and possible solutions (even if not originally framed in terms of collaboration) to triangulate our interview results. Relevant literature spans multiple research communities and publication venues, including machine learning, human-computer interaction, software engineering, systems, and various application domains (e.g., healthcare, finance), and does not always include obvious keywords; simply searching for machine-learning research yields a far too wide net. Hence, we decided against a systematic literature review and pursued a best effort approach that relied on keyword search for topics surfaced in the interviews, as well as backward and forward snowballing. Out of over 300 papers read, we identified 61 as possibly relevant and coded them with the same evolving codebook. The complete list can be found in Supplement D.

\myparagraph{Step 4: Validity check with interviewees.}  For checking fit and applicability as defined by Corbin and Strauss \cite{ia1C} and validating our findings, we went back to the interviewees after creating a full draft of this paper. We presented the interviewees both a summary and the full draft, including the supplementary material, along with questions prompting them to look for correctness and areas of agreement or disagreement (i.e., fit), and any insights gained from reading about experiences of the other companies, roles, or findings as a whole (i.e., applicability).
Ten interviewees responded with comments and all indicated general agreement, some explicitly reaffirmed some findings. We incorporated two minor suggested changes about details of two organizations.

\myparagraph{Threats to validity and credibility.} Our work exhibits the typical threats common and expected for this kind of qualitative research. Generalizations beyond the sampled participant distribution should be made with care; for example, we interviewed few managers, no dedicated data experts, and no clients. In several organizations, we were only able to interview a single person, giving us a one-sided perspective. Observations may be different in organizations in specific domains or geographic regions not well represented in our data. Self-selection of participants may influence results; for example developers in government-related projects more frequently declined interview requests. As described earlier, we followed standard practices for coding and memoing, but, as usual in qualitative research, we cannot entirely exclude biases introduced by us researchers.

\section{Diversity of Org. Structures}

\label{h.2fm47zvowv12}
Throughout our interviews, we found that the number and type of teams that participate in ML-enabled system development differs widely, as do their composition and responsibilities, power dynamics, and the formality of their collaborations, in line with findings by Aho et al. \cite{aho}. To illustrate these differences, we provide simplified descriptions of teams found in two organizations in Figure~\ref{fig:orgstructure}. We show teams and their members, as well as the artifacts for which they are \emph{responsible}, such as, who develops the \emph{model}, who builds a repeatable \emph{pipeline}, who operates the model (\emph{inference}), who is responsible for or owns the \emph{data}, and who is responsible for the final \emph{product}. A team often has multiple responsibilities and interfaces with other teams at multiple collaboration points. Where unambiguous, we refer to teams by their primary responsibility as \emph{product team} or \emph{model team}.

Organization 3 (Figure~\ref{fig:orgstructure}, top) develops an ML-enabled system for a government client. The product (health domain), including an ML model and multiple non-ML components, is developed by a single 8-person team. The team focuses on training a model first, before building a product around it. Software engineering and data science tasks are distributed within the team, where members cluster into groups with different responsibilities and roughly equal negotiation power. A single data scientist is part of this team, though they feel somewhat isolated. Data is sourced from public sources. The relationship between the client and development team is somewhat distant and formal. The product is delivered as a service, but the team only receives feedback when things go wrong.

Organization 7 (Figure~\ref{fig:orgstructure}, bottom) develops a product for in-house use (quality control for a production process). A small team is developing and using the product, but model development is delegated to an external team (different company) composed of four data scientists, of which two have some software engineering background. The product team interacts with the model team to define and revise model requirements based on product requirements. The product team  provides confidential proprietary data for training. The model team deploys the model and provides a ready-to-use inference API to the product team. The relationship between the teams crosses company boundaries and is rather distant and formal. The product team clearly has the power in negotiations between the teams.

These two organizations differed along many dimensions,  and we found no clear global patterns when looking across organizations. Nonetheless patterns did emerge when focusing on three specific collaboration aspects, as we will discuss in the next sections.

\section{Collaboration Point: Requirements and Planning}

\label{h.5yfe7wvf6ofp}
In an idealized top-down process, one would first solicit \emph{product}\emph{ requirements} and then plan and design the product by dividing work into components (ML and non-ML), deriving each \emph{component}\emph{'s}\emph{ requirements/specifications} from the product requirements. In this process, collaboration is needed for: (1) product team needs to negotiate \emph{product }requirements with clients and other stakeholders; (2) product team needs to plan and design product decomposition, negotiating with component teams the requirements for individual components; and (3) product project manager needs to plan and manage the work across teams in terms of budgeting, effort estimation, milestones, and work assignments.

\subsection{Common Development Trajectories}

\label{h.nuhso0i4hzev}
Few organizations, if any, follow an idealized top-down process, and it may not even be desirable, as we will discuss later. While we did not find any global patterns for organizational structures (Sec.~\ref{h.2fm47zvowv12}), there are indeed distinct patterns relating to how organizations elicit requirements and decompose their systems. Most importantly, we see differences in terms of the \emph{order} in which teams identify product and model requirements:

\textbf{Model-first trajectory:} 13 of the 28 organizations (3, 10, 14--17, 19, 20, 22, 23, 25--27) focus on building the model first, and build a product around the model later. In these organizations, product requirements are usually shaped by model capabilities after the (initial) model has been created, rather than being defined upfront. In organizations with separate model and product teams, the model team typically starts the project and the product team joins later with low negotiating power to build a product around the model.

\textbf{Product}\textbf{-first trajectory:} In 13 organizations (1, 4, 5, 7--9, 11--13, 18, 21, 24, 28), models are built later to support an existing product. In these cases, a product often already exists and product requirements are collected for how to extend the product with new ML-supported functionality. Here, the model requirements are derived from the product requirements and often include constraints on model qualities, such as latency, memory and explainability.

\textbf{Parallel trajectory:} Two organizations (2, 6) follow no clear temporal order; model and product teams work in parallel.

\subsection{Product and Model Requirements}

\label{h.6tv7b87udh5z}
We found a constant tension between product and model requirements in our interviews. Functional and nonfunctional product requirements set expectations for the entire product. Model requirements set goals and constraints for the model team, such as expected accuracy and latency, target domain, and available data.

\myparagraph{Product requirements require input from the model team (\faGroup, \faCalendar).} A common theme in the interviews is that it is difficult to elicit product requirements without a good understanding of ML capabilities, which almost always requires involving the model team and performing some initial modeling when eliciting product requirements. Regardless of whether product requirements or model requirements are elicited first, data scientists often mentioned being faced with \emph{unrealistic expectations} about model capabilities.

Participants that interact with clients to negotiate product requirements (which may involve members of the model team) indicate  that they need to educate  clients about capabilities of ML techniques to \emph{set correct expectations} (P3a, P6a, P6b, P7b, P9a, P10a, P15c, P19b, P22b, P24a). This need to educate customers about ML capabilities has also been raised  in the literature \cite{JnkY,aFTn,We4r,FdO8,bU0N,kckt,aho}.

For many organizations, especially in \emph{product-first trajectories}, the model team indicates similar challenges when interacting with the product team. If the product team does not involve the model team in negotiating product requirements, the product team may not identify what data is needed for building the model, and may commit to unrealistic requirements. For example, P26a shared \emph{``For this project, [the project manager] wanted to claim that we have no false positives and I was like, that's not gonna work.'' } Members of the model team often report lack of ML literacy in members of the product team and project managers (P1b, P4a, P7a, P12a, P26a, P27a) and a lack of involvement (e.g., P7b: ``\emph{The [product team] decided what type of data would make sense. I had no say on that.''}). Usually the product team cannot identify product requirements alone, instead product and model teams need to interact to explore what is achievable.

In organizations with a \emph{model-first trajectory}, members of the model team sometimes engage directly with clients (and also report having to educate them about ML capabilities). However, when requirements elicitation is left to the model team, members tend to focus on requirements relevant for the model, but neglect requirements for the product, such as expectations for usability, e.g., P3c's customers ``\emph{were kind of happy with the results, but }\emph{weren't happy with the overall look and feel or }\emph{how the system worked.''} Several research papers similarly identified how the goals of data scientists diverge from product goals if product requirements are not obvious at modeling time, leading to inefficient development, worse products, or constant renegotiation of requirements, especially~\cite{passi,ZfhU,yang}.

\myparagraph{Model development with unclear model requirements is common (\faFileText).} Participants from model teams frequently explain how they are expected to work independently, but are given sparse model requirements. They try to infer intentions behind them, but are constrained by having limited understanding of the product that the model will eventually support (P3a, P3b, P16b, P17b, P19a). Model teams often start with vague goals and model requirements evolve over time as product teams or clients refine their expectations in response to provided models (P3b, P7a, P9a, P5b, P19b, P21a). Especially in organizations following the \emph{model-first trajectory}, model teams may receive some data and a goal to predict something with high accuracy, but no further context, e.g., P3a shared \emph{``there isn't always an actual spec of exactly what data they have, what data they think they're going to have and what they want the model to do.''} Several papers similarly report projects starting with vague model goals \cite{vhGB,5Nhu,CY3n,kuwajima}.

Even in organizations following a \emph{product-first trajectory}, product requirements are often not translated into clear model requirements. For example, participant P17b reports how the model team was not clear about the model's intended target domain, thus could not decide what data was considered in scope. As a consequence, model teams usually cannot focus just on their component, but have to understand the entire product to identify model requirements in the context of the product (P3a, P10a, P13a, P17a, P17b, P19b, P20b, P23a), requiring interactions with the product team or even bypassing the product team to talk directly to clients. The difficulty of providing clear requirements for an ML model has also been raised in the literature \cite{JnkY,ExZ5,q6tY,5Nhu,8e8c,7dgT}, partially arguing that uncertainty makes it difficult to specify model requirements upfront \cite{FdO8,ctC1,bU0N,kuwajima,aho}. Ashemore et al. report mapping product requirements to model requirements as an open challenge \cite{Pd6Q}. \looseness=-1

\myparagraph{Provided model requirements rarely go beyond accuracy and data security (\faGears, \faFileText).} Requirements given to model teams primarily relate to some notion of accuracy. Beyond accuracy, requirements for data security and privacy are common, typically imposed by the data owner or by legal requirements (P5a, P7a, P9a, P13a, P14a, P18a, P20a-b, P21a-b, P22a, P23a, P24a, P25a, P26a). Literature also frequently discusses how privacy requirements impact and restrict ML work \cite{Yuwe,JN0G,q6tY,K8oj,Mfyd,efZw}.

We rarely heard of any qualities other than accuracy. Some participants report that ignoring qualities such as latency or scalability has resulted in integration and operation problems (P3c, P11a). In a few cases requirements for inference latency were provided (P1a, P6a, P14a) and in one case hardware resources provided constraints on memory usage (P14a), but no other qualities such as training latency, model size, fairness, or explainability were required that could be important for product integration and deployment.

When prompted, very few of our interviewees report considerations for fairness either at the product or the model level. Only two participants from model teams (P14a, P22a) reported receiving fairness requirements, whereas many others explicitly mentioned that fairness is not a concern for them yet (P4a, P5b, P6b, P11a, P15c, P20a, P21b, P25a, P26a). The lack of fairness and explainability requirements is in stark contrast to the emphasis that these qualities receive in the literature \cite[e.g.,][]{7LH3,sWPF,Emzk,Yuwe,7dgT,9t4Z,ynnp,uoFv,TIPJ,gC0G}.

\myparagraph{Recommendations.} Our observations suggest that involving data scientists early when soliciting product requirements is important (\faGroup) and that pursuing a model-first trajectory entirely without considering product requirements is problematic (\faCalendar). Conversely, model requirements are rarely specific enough to allow data scientists to work in isolation without knowing the broader context of the system and interaction with the product team should ideally be planned as part of the process. Requirements form a key collaboration point between product and model teams, which should be emphasized even in more distant collaboration styles (e.g., outsourced model development). The few organizations that use the \emph{parallel trajectory} report fewer problems by involving data scientists in negotiating product requirements to discard unrealistic ones early on (P6b). Vogelsang and Borg also provide similar recommendations to consult data scientists from the beginning to help elicit requirements \cite{aFTn}. While many papers place emphasis on clearly defining ML use cases and scope \cite{JnkY,kckt,Dzd7}, several others mention how collaboration of technical and non-technical stakeholders such as domain experts helps \cite{9t4Z,8e8c,bU0N,passi}.

ML literacy for customers and product teams appears to be important (\faGroup). P22a and P19a suggested conducting technical ML training sessions to educate clients; similar training is also useful for members of product teams. Several papers argue for similar training for non-technical users of ML products \cite{FdO8,9t4Z,aFTn}. \looseness=-1

Most organizations elicit requirements only rather informally and rarely have good documentation, especially when it comes to model requirements. It seems beneficial to adopt more formal requirements documentation for product and model (\faFileText), as several participants reported that it fosters shared understanding at this collaboration point (P11a, P13a, P19b, P22a, P22c, P24a, P25a, P26a). Checklists could help to cover a broader range of model quality requirements, such as training latency, fairness, and explainability. Formalisms such as model cards \cite{uiNH} and FactSheets \cite{factsheet} could be used as a starting point for documenting model requirements.

\subsection{Project Planning}

\label{h.36mvexeh3lz}
\myparagraph{ML uncertainty makes effort estimation difficult (\faGroup).} Irrespective of trajectory, 19 participants (P3a, P4a, P7a-b, P8a, P14b, P15b-c, P16a, P17a, P18a, P19a-b, P20a, P22a-c, P23a, P25a) mentioned that the uncertainty associated with ML components makes it difficult to estimate the timeline for developing an ML component and by extension the product. Model development is typically seen as a science-like activity, where iterative experimentation and exploration is needed to identify whether and how a problem can be solved, rather than as an engineering activity that follows a somewhat predictable process. This science-like nature makes it difficult for the model team to set expectations or contracts with clients or the product team regarding effort, cost, or accuracy. While data scientists find effort estimation difficult, lack of ML literacy in managers makes it worse (P15b, P16a, P19b, P20a, P22b). Teams report deploying subpar models when running out of time (P3a, P15b, P19a), or postponing or even canceling deployments (P25a). These findings align with literature mentioning difficulties associated with effort estimation for ML tasks  \cite{9g1T,B5qD,bU0N,aho} and planning projects in a structured manner with diverse methodologies, with diverse trajectories, and without practical guidance \cite{We4r,B5qD,bU0N,aho}.  \looseness=-1

Generally, participants frequently report that synchronization between teams is challenging because of different team pace, different development processes, and tangled responsibilities (P2a, P11a, P12a, P14-b, P15b-c, P19a; see also Sec. \ref{h.4ifgk9l6syc2}).

\myparagraph{Recommendations.} Participants suggested several mitigation strategies: keeping extra buffer times and adding additional time-boxes for R\&D in initial phases (P8a, P19a, P22b-c, P23a; \faCalendar), continuously involving clients in every phase so that they can understand the progression of the project and be aware of potential missed deadlines (P6b, P7a, P22a, P23a; \faGroup). From the interviews, we also observe the benefits of managers who understand both software engineering and machine learning and can align product and model teams toward common goals (P2a, P6a, P8a, P28a; \faGroup).

\section{Collaboration Point: Training Data}

\label{h.drnfz16w2g29}
Data is essential for machine learning, but disagreements and frustrations around training data were the most common collaboration challenges mentioned in our interviews. In most organizations, the team that is responsible for building the model is not the team that collects, owns, and understands the data, making data a key collaboration point between teams in ML-enabled systems development.

\subsection{Common Organizational Structures}

\label{h.13ovdl1d0cn2}
We observed three patterns around data that influence collaboration challenges from the perspective of the model team:

\begin{wrapfigure}{r}{0.15\textwidth}
  \begin{center}
    \includegraphics[trim=2.5cm 1.5cm 0.5cm 1.5cm, width=0.15\textwidth]{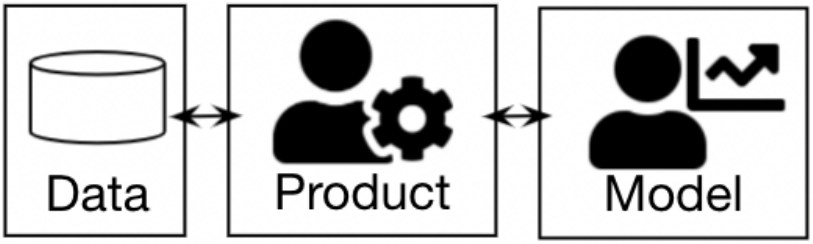}
  \end{center}
\end{wrapfigure}
\textbf{Provided data}\textbf{:} The product team has the responsibility of providing data to the model team (org. 6--8, 13, 18, 21, 23). The product team is the initial point of contact for all data-related questions from the model team. The product team may own the data or acquire it from a separate data team (internal or external). Coordination regarding data tends to be distant and formal, and the product team tends to hold more negotiation power.

\begin{wrapfigure}{r}{0.15\textwidth}
  \begin{center}
    \includegraphics[trim=2.5cm 1.5cm 0.5cm 1.5cm, width=0.15\textwidth]{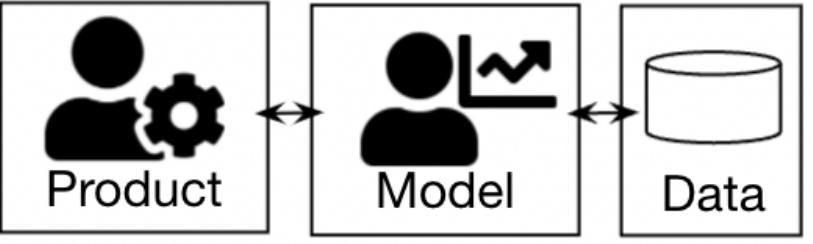}
  \end{center}
\end{wrapfigure}
\textbf{External data}\textbf{: }The product team does not have direct responsibility for providing data, but instead, the model team relies on external data providers. Commonly, the model team (i) uses publicly available resources (e.g., academic datasets, org. 2--4, 6, 19) or (ii) hires a third party for collecting or labeling data (org. 9, 15--17, 22, 23). In the former case, the model team has little to no negotiation power over data; in the latter, it can set expectations.

\begin{wrapfigure}{r}{0.15\textwidth}
  \begin{center}
    \includegraphics[trim=2.5cm 1.5cm 0.5cm 2.5cm, width=0.15\textwidth]{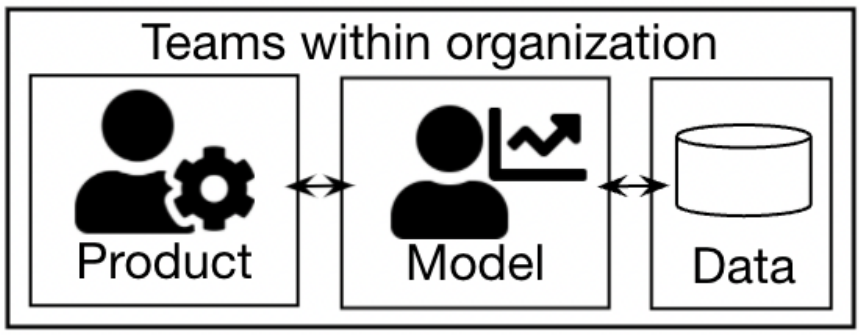}
  \end{center}
\end{wrapfigure}
\textbf{In-house data}\textbf{: }Product, model, and data teams are all part of the same organization and the model team relies on internal data from that organization (org. 1, 5, 9--12, 14, 20, 24--28). In these cases, both product and model teams often find it challenging to negotiate access to internal data due to differing priorities, internal politics, permissions, and security constraints.

\subsection{Negotiating Data Quality and Quantity}

\label{h.ng5qs5o69kjs}
Disagreements and frustrations around training data were the most common collaboration challenges in our interviews. In almost every project, data scientists were unsatisfied with the quality and quantity of data they received at this collaboration point, in line with a recent survey showing data availability and management to be the top-ranked challenge in building ML-enabled systems \cite{tPQ5}.

\myparagraph{Provided and public data is often inadequate (\faFileText, \faGroup).} In organizations where data is provided by the product team, the model team commonly states that it is difficult to get sufficient data (P7a, P8a, P13a, P22a, P22c). The data that they receive is often of low quality, requiring significant investment in data cleaning. Similar to the requirements challenges discussed earlier, they often state that the product team has little knowledge or intuition for the amount and quality of data needed. For example, participant P13a stated that they were given a spreadsheet with only 50 rows to build a model and P7a reported having to spend a lot of time convincing the product team of the importance of data quality. This aligns with past observations that software engineers often have little appreciation for data quality concerns \cite{WVIq,7wRo,JnkY,CY3n,muiruri} and that training data is often insufficient and incomplete \cite{vhGB,x4oR,CY3n,efZw,bU0N,Mfyd,Dzd7}.

When the model team uses public data sources, its members also have little influence over data quality and quantity and report significant effort for cleaning low quality and noisy data (P2a, P3a, P4a, P3c, P6b, P19b, P23a). Papers have similarly questioned the representativeness and trustworthiness of public training data \cite{ynnp,aFTn,5guS} as \emph{``nobody gets paid to maintain such data''} [104].

\emph{Training-serving skew} is a common challenge when training data is provided to the model team: models show promising results, but do not generalize to production data because it differs from provided training data (P4a, P8a, P13a, P15a, P15c, P21a, P22c, P23a) \cite{9g1T,Lmxi,kckt,JN0G,q6tY,ynnp,CY3n,7QTl,Mfyd,Jmep,WVIq}. Our interviews show that this skew often originates from inadequate training data combined with unclear information about production data, and therefore no chance to evaluate whether the training data is representative of production data.

\myparagraph{Data understanding and access to domain experts is a bottleneck (\faFileText, \faCalendar). }Existing data documentation (e.g, data item definitions, semantics, schema) is almost never sufficient for model teams to understand the data (also mentioned in a prior study \cite{hzdi}). In the absence of clear documentation, team members often collect information and keep track of unwritten details in their heads (P5a), known as institutional or tribal knowledge \cite{gC0G,tPQ5}. Data understanding and debugging often involve members from different teams and thus cause challenges at this collaboration point.

Model teams receiving data from the product team report struggling with data understanding and having a difficult time getting help from the product team (or the data team that the product team works with) (P8a, P7b, P13a). As the model team does not have direct communication with the data team, data understanding issues often cannot be resolved effectively. For example, P13a reports \emph{``Ideally, for us it would be so good to spend maybe a week or two with one person continuously trying to understand the data. It's one of the biggest problems actually, because even if you have the person, if you're not in contact all the time, then you misinterpreted some things and you build on it.''} The low negotiation power of the model team in these organizations hinders access to domain experts.

Model teams using public data similarly struggle with data understanding and getting help (P3a, P4a, P19a), relying on sparse data documentation or trying to reach any experts on the data. \looseness=-1

For in-house projects, in several organizations the model team relies on data in shared databases (org. 5, 11, 26, 27, 28), collected by instrumenting a production system, but shared by multiple teams. Several teams shared problems with evolving and often poorly documented data sources, as participant P5a illustrates \emph{``[data rows] can have 4,000 features, 10,000 features. And no one really cares. They just dump features there. [...] I just cannot track 10,000 features.''} Model teams face challenges in understanding data and identifying a team that can help (P5a, P25a, P20b, P27a), a problem also reported in a prior study about data scientists at Microsoft \cite{JnkY}.

Challenges in understanding data and needing domain experts are also frequently mentioned in the literature  \cite{JnkY,CY3n,hzdi,6aXp,K8oj,gC0G,WVIq,muiruri}, as is the danger of building models with insufficient understanding of the data \cite{aFTn,5guS}. Although we are not aware of literature discussing the challenges of accessing domain experts, papers have shown that even when data scientists have access, effective knowledge transfer is challenging \cite{yL64,42j1}.

\myparagraph{Ambiguity when hiring a data team (\faFileText).} When the model team hires an external data team for collecting or labelling data (org. 9, 15, 16, 17, 22, 23), the model team has much more negotiation power over setting data quality and quantity expectations (though Kim et al. report that model teams may have difficulty getting buy-in from the product team for hiring a data team in the first place \cite{JnkY}). Our interviews did not surface the same frustrations as with provided data and public data, but instead participants from these organizations reported \emph{communication vagueness} and \emph{hidden assumptions} as key challenges at this collaboration point (P9a, P15a, P15c, P16a, P17b, P22a, P22c, P23a). For example, P9a related how different labelling companies given the same specification widely disagreed on labels, when the specification was not clear enough.

We found that expectations between model and data teams are often communicated verbally without clear documentation. As a result, the data team often does not have sufficient context to understand what data is needed.  For example, participant P17b states \emph{``Data collectors can't understand the data requirements all the time. Because, when a questionnaire [for data collection] is designed, the overview of the project is not always described to them. Even if we describe it, they can't always catch it.'' }Reports about low quality data from hired data teams have been also discussed in the literature~\cite{aFTn,Pd6Q,q6tY,bU0N,efZw,WVIq}.

\myparagraph{Need to handle evolving data (\faGears, \faGroup).} In most projects, models need to be regularly retrained with more data or adapted to changes in the environment (e.g., data drift) \cite{hpbm,q6tY,WVIq}, which is a challenge for many model teams (P3a, P3c, P5a, P7a-b, P11a, P15c, P18a, P19b, P22a). When product teams provide the data, they often have a static view and provide only a single snapshot of data rather than preparing for updates, and model teams with their limited negotiation power have a difficult time fostering a more dynamic mindset (P7a-b, P15c, P18a, P22a), as expressed by participant P15c: \emph{``People don't understand that for a machine learning project, data has to be provided constantly.''} It can be challenging for a model team to convince the product team to invest in continuous model maintenance and evolution (P7a, P15c) \cite{hzdi}.

Conversely, if data is provided continuously (most commonly with public data sources, in-house sources, and own data teams), model teams struggle with ensuring consistency over time. Data sources can suddenly change without announcement (e.g., changes to schema, distributions, semantics), surprising model teams that make but do not check assumptions about the data (P3a, P3c,  P19b). For example, participants P5a and P11a report similar challenges with in-house data, where their low negotiation power does not allow them to set quality expectations, but they  face undesired and unannounced changes in data sources made by other teams. Most organizations do not have a monitoring infrastructure to detect changes in data quality or quantity, as we will discuss in Sec.~\ref{h.ytgc9mej2ee2}.

\myparagraph{In-house priorities and security concerns often obstruct data access (\faCalendar).} In in-house projects, we frequently heard about the product or model team struggling to work with another team within the same organization that owns the data. Often, these in-house projects are local initiatives (e.g., logistics optimization) with more or less buy-in from management and without buy-in from other teams that have their own priorities; sometimes other teams explicitly question the business value of the product. The interviewed model teams usually have little negotiation power to request data (especially if it involves collecting additional data) and almost never get an agreement to continuously receive data in a certain format, quality, or quantity (P5a, P10a, P11a, P20a-b, P27a) (also observed in studies at Microsoft, ING and other organizations  \cite{JnkY,5guS,muiruri}). For example, P10a shared \emph{``we wanted to ask the data warehouse team to [provide data], and it was really hard to get resources. They wouldn't do that because it was hard to measure the impact [our in-house project] had on the bottom line of the business.''} Model teams in these settings tend to work with whatever data they can get eventually.

Security and privacy concerns can also limit access to data  (P7a, P7b, P21a-b, P22a, P24a) \cite{CY3n,hzdi,q6tY,Mfyd,muiruri}, especially when data is owned by a team in a different organization, causing frustration, lengthy negotiations, and sometimes expensive data-handling restrictions (e.g., no use of cloud resources) for model teams.

\myparagraph{Recommendations.} Data quality and quantity is important to model teams, yet they often find themselves in a position of low negotiation power, leading to frustration and collaboration inefficiencies. Model teams that have the freedom to set expectations and hire their own data teams are noticeably more satisfied. When planning the entire product, it seems important to pay special attention to this collaboration point, and budget for data collection, access to domain experts, or even a dedicated data team (\faCalendar). Explicitly planning to provide substantial access to domain experts early in the project was suggested as important (P25a).

We found it surprising that despite the importance of this collaboration point there is little written agreement on expectations and often limited documentation (\faFileText), even when hiring a dedicated data team---in stark contrast to more established contracts for traditional software components. Not all organizations allow the more agile, constant close collaboration between model and data teams that some suggest \cite{CY3n,7QTl}. With a more formal or distant relationship (e.g., across organizations, teams without buy-in), it seems beneficial to adopt a \emph{more formal contract}, specifying data quantity and quality expectations, which are well researched in the database literature \cite{jt5V} and have been repeatedly discussed in the context of ML-enabled systems \cite{yL64,JnkY,hzdi,efZw,Mfyd}. This has also been framed as \emph{data requirements} in the software engineering literature \cite{aFTn,kckt,vhGB}. When working with a dedicated data team, participants suggested to invest in making expectations very clear, for example, by providing precise specifications and guidelines (P9a, P6b, P28a), running training sessions for the data collectors and annotators (P17b, P22c), and measuring inter-rater agreement (P6b).

Automated checks are also important as data evolves (\faGears). For example, participant P13a mentioned proactively setting up data monitoring to detect problems (e.g., schema violations, distribution shifts) at this collaboration point; a practice suggested also in the literature \cite{kckt,9t4Z,CY3n,7QTl,Mfyd,WVIq,lewis2021} and supported by recent tooling  \cite[e.g.,][]{ubVS,7QTl,kj2C}. The risks regarding possible unnoticed changes to data make it important to consider \emph{data validation }and \emph{monitoring infrastructure} as a key feature of the product early on (\faGears, \faCalendar), as also emphasized by several participants (P5a, P25a, P26a, P28a).

\section{Collaboration Point: Product-Model Integration}

\label{h.6n0xem26clpb}
As discussed earlier, to build an ML-enabled system both ML components and traditional non-ML components need to be integrated and deployed, requiring data scientists and software engineers to work together, typically across multiple teams. We found many conflicts at this collaboration point, stemming from unclear processes and responsibilities, as well as differing practices and expectations.

\subsection{Common Organizational Structures}

\label{h.mz7k5sr2sgu1}
We saw large differences among organizations in how engineering responsibilities were assigned, most visible in how responsibility for \emph{model deployment and operation} is assigned, which typically involves significant engineering effort for building reproducible pipelines, API design, or cloud deployment, often with MLOps technologies. We found the following patterns:

\begin{wrapfigure}{r}{0.21\textwidth}
  \begin{center}
    \includegraphics[trim=2.5cm 1.75cm 0cm 2cm, scale=.23]{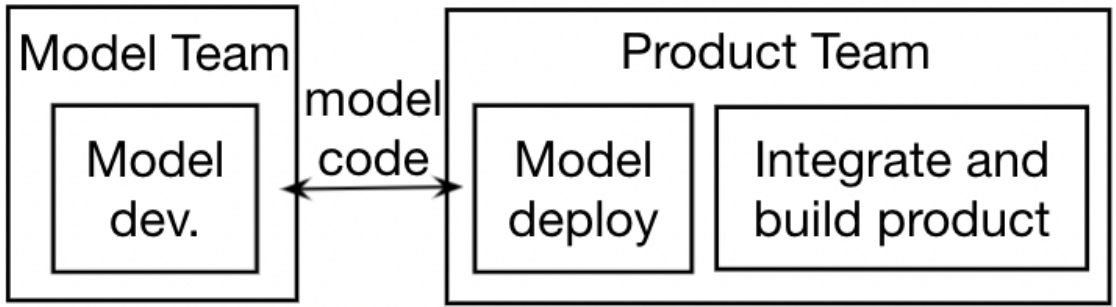}
  \end{center}
\end{wrapfigure}
\textbf{Shared model code:} In some organizations (2, 6, 23, 25), the model team is responsible only for model development and delivers training code (e.g., in a notebook) or model files to the product team; the product team takes responsibility for deployment and operation of the model, possibly rewriting the training code as a pipeline. Here, the model team has little or no engineering responsibilities.

\begin{wrapfigure}{r}{0.19\textwidth}
  \begin{center}
    \includegraphics[trim=2.5cm 1.75cm 0cm 2cm, scale=.23]{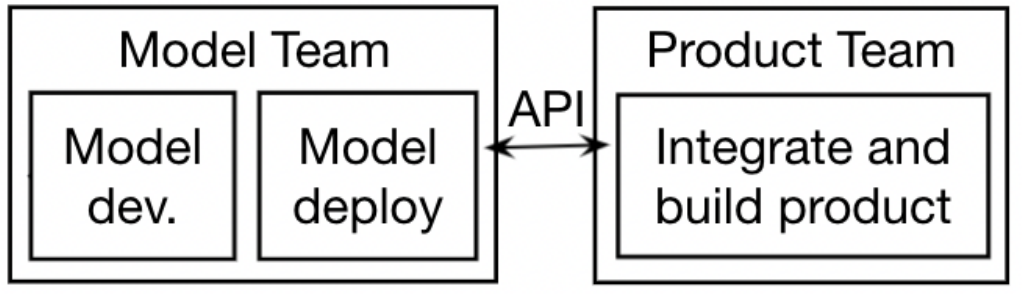}
  \end{center}
\end{wrapfigure}
\textbf{Model as API: }In most organizations (18 out of 28), the model team is responsible for developing and deploying the model. Hence, the model team requires substantial engineering skills in addition to data science expertise. 
Here, some model teams are mostly composed of data scientists with little engineering capabilities (org. 7, 13, 17, 22, 26), some consist mostly of software engineers who have picked up some data science knowledge (org. 4, 15, 16, 18, 19, 21, 24), and others have mixed team members (org. 1, 9, 11, 12, 14, 28). These model teams typically provide an API to the product team, or release individual model predictions (e.g., shared files, email; org. 17, 19, 22) or install models directly on servers (org. 4, 9, 12).

\begin{wrapfigure}{r}{0.17\textwidth}
  \begin{center}
    \includegraphics[trim=2.5cm 1.75cm 0cm 2cm, scale=.23]{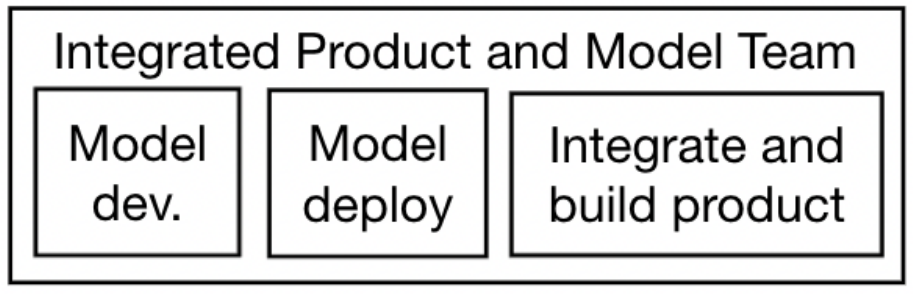}
  \end{center}
\end{wrapfigure}
\textbf{All-in-one:} If only few people work on model and product, sometimes a single team (or even a single person) shares all responsibilities (org. 3, 5, 10, 20, 27). 
It can be a small team with only data scientists (org. 10, 20, 27
) or mixed teams with data scientists and software engineers (org. 3, 5).

We also observed two outliers
: One startup (org. 8) had a distinct model deployment team, allowing the model team to focus on data science without much engineering responsibility. In one large organization (org. 28), an engineering-focused model team (model as API) was supported by a dedicated \emph{research team} focused on data-science research with fewer engineering responsibilities.

\subsection{Responsibility and Culture Clashes}

\label{h.4ifgk9l6syc2}
Interdisciplinary collaboration is challenging (cf. Sec.~\ref{h.kpwvxerzqjbu}). We observed many conflicts between data science and software engineering culture, made worse by unclear responsibilities and boundaries.

\myparagraph{Team responsibilities often do not match capabilities and preferences (\faGears).} When the model team has responsibilities requiring substantial engineering work, we observed some dissatisfaction when its members were assigned undesired responsibilities. Data scientists preferred engineering support rather than needing to do everything themselves (P7a-b, 13a), but can find it hard to convince management to hire engineers (P10a, P20a, P20b). For example P10a describes \emph{``I was 
struggling to change the mindset of the team lead, convincing him to hire an engineer...
I just didn't want this to be my main responsibility.'' }Especially in small teams, data scientists report struggling with the complexity of the typical ML infrastructure (P7b, P9a, P14a, P26a, P28a).
\looseness=-1

In contrast, when deployment is the responsibility of software engineers in the product team or of dedicated engineers in \emph{all-in-one} teams, some of those engineers report problems integrating the models due to insufficient knowledge on model context or domain, and the model code not being packaged well for deployment (P20b, P23a, P27a). In several organizations, we heard about software engineers performing ML tasks without having enough ML understanding (P5a, P15b-c, P16b, 18b, 19b, 20b). Mirroring observations from past research \cite{5Nhu}, P5a reports \emph{``there are people who are ML engineers at [company] , but they don't really understand ML. They were actually software engineers... 
they don't understand [overfitting, underfitting, ...]. They just copy-paste code.'' 
}

\myparagraph{Siloing data scientists fosters integration problems (\faGroup, \faCalendar).} We observed data scientists often working in isolation---known as \emph{siloing}---in all types of organizational structures, even within single small teams (see Sec.~\ref{h.2fm47zvowv12}) and within engineering-focused teams. In such settings, data scientists often work in isolation with weak requirements (cf. Sec. \ref{h.6tv7b87udh5z}) without understanding the larger context, seriously engaging with others only during integration (P3a, P3c, P6a, P7b, P11a, P13a, P15b, P25a) \cite{K8oj}, where problems may surface. For example, participant P11a reported a problem where product and model teams had different assumptions about the expected inputs and the issue could only be identified after a lot of back and forth between teams at a late stage in the project.

\myparagraph{Technical jargon challenges communication (\faGroup).} Participants frequently described communication issues arising from differing terminology used by members from different backgrounds (P1a-b, P2a, P3a, P5b, P8a, P12a, P14a-b, P16a, P17a-b, P18a-b, P20a, P22b, P23a), leading to ambiguity,  misunderstandings, and inconsistent assumptions (on top of communication challenges with domain experts) \cite{hzdi,evYm,8e8c,aho}. 
P1b reports, \emph{``There are a lot of conversations in which disambiguation becomes necessary. We often use different kinds of words that might be ambiguous.'' }For example, data scientists may refer to prediction accuracy as \emph{performance}, a term many software engineers associate with response time. These challenges can be observed more frequently between teams, but they even occur within a 
team with members from different backgrounds (P3a-c, P20a).

\myparagraph{Code quality, documentation, and versioning expectations differ widely and cause conflicts (\faGroup, \faGears).} Many participants reported conflicts around development practices between data scientists and software engineers 
during integration and deployment. Participants report poor practices that may also be observed in traditional software projects; but particularly software engineers expressed frustration in interviews that data scientists do not follow the same development practices or have the same quality standards when it comes to writing code. Reported problems relate to poor code quality (P1b, P2a, P3b, P5a, P6a-b, P10a, P11a, P14a, P15b-c, P17a, P18a, P19a, P20a-b, P26a) \cite{MAdB,4sUH,aGd5,9g1T,bU0N,5guS,7sej}, insufficient documentation (P5a-b, P6a-b, P10a, P15c, P26a)  \cite{hzdi,7LH3,uiNH, factsheet}, and not extending version control to data and models (P3c, P7a, P10a, P14a, P20b). In two shared-model-code organizations, participants report having to rewrite code from the data scientists (P2a, P6a-b). Missing documentation for ML code and models is considered the cause for different assumptions that lead to incompatibility between ML and non-ML components (P10a) and for losing knowledge and even the model when faced with turnover (P6a-b). Recent papers similarly hold poor documentation responsible for team decisions becoming invisible and inadvertently causing hidden assumptions \cite{efZw,5guS,evYm,gC0G,7LH3,hzdi}. Hopkins and Booth called model and data versioning in small companies as desired but ``elusive'' \cite{gC0G}.

\myparagraph{Recommendations.} Many conflicts relate to boundaries of responsibility (especially for engineering responsibilities) and to different expectations by team members with different backgrounds. Better teams tend to define processes, responsibilities, and boundaries more carefully (\faCalendar), document APIs at collaboration points between teams (\faFileText), and recruit dedicated engineering support for model deployment (\faGears), but also establish a team culture with mutual understanding and exchange (\faGroup). Big tech companies usually have more established processes and clearer responsibility assignments than smaller organizations and startups that often follow ad-hoc processes or figure out responsibilities as they go.

The need for engineering skills for ML projects has frequently been discussed \cite{tPQ5,ZfhU,XeyT,aGd5,UjZF,Jmep,yang}, but our interviewees differ widely in whether all data scientists should have substantial engineering responsibilities or whether engineers should support data scientists so that they can focus on their core expertise (\faGears). Especially interviewees from big tech emphasized that they expect engineering skills from all data science hires (P28a). Others emphasized that recruiting software engineers and operations staff with basic data-science knowledge can help at many communication and integration tasks, such as converting experimental ML code for deployment (P2a, P3b), fostering communication (P3c, P25a), and monitoring models in production (P5b). Generally, \emph{siloing data scientists is widely recognized as problematic} and many interviewees suggest practices for improving communication (\faGroup), such as training sessions for establishing common terminology (P11a, P17a, P22a, P22c, P23a), weekly all-hands meetings to present all tasks and synchronize (P2a, P3c, P6b, P11a), and proactive communication to broadcast upcoming changes in data or infrastructure (P11a, P14a, P14b). This mirrors suggestions to invest in interdisciplinary training \cite{JnkY,tPQ5,evYm,ctC1,eIMb,yang} and proactive communication \cite{7wRo}.

\subsection{Quality Assurance for Model and Product}

\label{h.ytgc9mej2ee2}
During development and integration, questions of responsibility for quality assurance frequently arise, often requiring coordination and collaboration between multiple teams. This includes evaluating components individually (including the model) as well as their integration and the whole system, often including evaluating and monitoring the system \emph{online} (in production).

\myparagraph{Model adequacy goals are difficult to establish (\faFileText, \faGroup).} Off\-line accuracy evaluation of models is almost always performed by the model team responsible for building the model, though often they have difficulty deciding locally when the model is good enough (P1a, P3a, P5a, P6a, P7a, P15b, P16b, P23a) \cite{FdO8,5guS}. As discussed in Sec. \ref{h.5yfe7wvf6ofp} and Sec. \ref{h.drnfz16w2g29}, model team members often receive little guidance on model adequacy criteria and are unsure about the actual distribution of production data. They also voice concerns about establishing ground truth, for example, needing to support data for different clients, and hence not being able to establish (offline) measures for model quality (P1b, P16b, P18a, P28a). As quality requirements beyond accuracy are rarely provided for models, model teams usually do not feel responsible for testing latency, memory consumption, or fairness (P2a, P3c, P4a, P5a, P6b, P7a, P14a, P15b, P20b). Whereas literature discussed challenges in measuring business impact of a model \cite{JnkY,Pd6Q,2kVg,efZw} and balancing business goals with model goals \cite{passi}, interviewed data scientists were concerned about this only with regards to convincing clients, managers or product teams to provide resources (P7a-b, P10a, P26a, P27a).

\myparagraph{Limited confidence without transparent model evaluation (\faFileText).} Participants in several organizations report that model teams do not prioritize model evaluation and have no systematic evaluation strategy (especially if they do not have established adequacy criteria they try to meet), performing occasional ``ad-hoc inspections'' instead (P2a, P15b, P16b, P18b, P19b, P20b, P21b, P22a, P22b). Without transparency about their test processes and test results, other teams voiced reduced confidence in the model, leading to skepticism to adopt the model (P7a, P10a, P21b, P22a).

\myparagraph{Unclear responsibilities for system testing (\faCalendar).} Teams often struggle with testing the entire product after integrating ML and non-ML components. Model teams frequently explicitly mentioned that they assume no responsibility for product quality (including integration testing and testing in production) and have not been involved in planning for system testing, but that their responsibilities end with delivering a model evaluated for accuracy (P3a, P14a, P15b, P25a, P26a). However, in several organizations, product teams also did not plan for testing the entire system with the model(s) and, at most, conducted system testing in an ad-hoc way (P2a, P6a, P16a, P18a, P22a). Recent literature has reported a similar lack of focus on system testing in product teams \cite{7LH3,6aXp}, mirroring also a focus in academic research on testing models rather than testing the entire system \cite{Pd6Q,i8ow}. Interestingly, some established software development organizations delegated testing to an existing separate quality assurance team with no process or experience testing ML products (P2a, P8a, P16a, P18b, P19a).

\myparagraph{Planning for online testing and monitoring is rare (\faCalendar, \faGears, \faGroup).} Due to possible training-serving skew and data drift, literature emphasizes the need for online evaluation \cite{hpbm,Kol9,Lmxi,kj2C,aFTn,UjZF,I17M,2kVg,Pd6Q,sAxw,6aXp,FdO8,aGd5,muiruri}. With collected telemetry, one can usually approximate both product and model quality, monitor updates, and experiment in production \cite{2kVg}. Online testing usually requires coordination among multiple teams responsible for product, model, and operation. We observed that most organizations do \emph{not} perform monitoring or online testing, as it is considered difficult, in addition to lack of standard process, automation, or even test awareness (P2a, P3a, P3b, P4a, P6b, P7a, P10a, P15b, P16b, P18b, P19b, 25a, P27a). Only 11 out of 28 organizations collected any telemetry; it is most established in big tech organizations. When to retrain models is often decided based on intuition or manual inspection, though many aspire to more automation (P1a, P3a, P3c, P5a, P10a, P22a, P25a, P27a). Responsibilities around online evaluation are often neither planned nor assigned upfront as part of the project.

Most model teams are aware of possible data drift, but many do not have any monitoring infrastructure for detecting and managing drift in production. If telemetry is collected, it is the responsibility of the product or operations team and it is not always accessible to the model team. Four participants report that they rely on manual feedback about problems from the product team (P1a, P3a, P4a, P10a). At the same time, others report that product and operation teams do not necessarily have sufficient data science knowledge to provide meaningful feedback (P3a, P3b, P5b, P18b, P22a) \cite{gWib}.

\myparagraph{Recommendations.} Quality assurance involves multiple teams and benefits from explicit planning and making it a high priority (\faCalendar). While the product team should likely take responsibility for product quality and system testing, such testing often involves building monitoring and experimentation infrastructure (\faGears), which requires planning and coordination with teams responsible for model development, deployment, and operation (if separate) to identify the right measures. Model teams benefit from receiving feedback on their model from production systems, but such support needs to be planned explicitly, with corresponding engineering effort assigned and budgeted, even in organizations following a model-first trajectory. We suspect that education about benefits of testing in production and common infrastructure (often under the label Dev\-Ops/ML\-Ops \cite{5hJ9}) can increase buy-in from all involved teams (\faGroup). Organizations that have established monitoring and experimentation infrastructure strongly endorse it (P5a, P25a, P26a, P28a). \looseness=-1

Defining clear quality requirements for model and product can help all teams to focus their quality assurance activities (cf. Sec. \ref{h.5yfe7wvf6ofp}; \faFileText). Even when it is challenging to define adequacy criteria upfront, teams can together develop a quality assurance plan for model and product. Participants and literature emphasized the importance of human feedback to evaluate model predictions (P11a, P14a) \cite{sAxw}, which requires planning to collect such feedback (\faCalendar). System and usability testing may similarly require planning for user studies with prototypes and shadow deployment \cite{ynnp,9t4Z,kckt}.

\section{Discussion and Conclusions}

\label{h.jdytnrpd8obn}
Through our interviews we identified three central collaboration points where organizations building ML-enabled systems face substantial challenges: (1) requirements and project planning, (2) training data, and (3) product-model integration. Other collaboration points surfaced, but were mentioned far less frequently (e.g., interaction with legal experts and operators), did not relate to problems between multiple disciplines (e.g., data scientists documenting their work for other data scientists), or mirrored conventional collaboration in software projects (e.g., many interviewees wanted to talk about unstable ML libraries and challenges interacting with teams building and maintaining such libraries, though the challenges largely mirrored those of library evolution generally \cite{4JFg,1i9g}).

Data scientists and software engineers are certainly not the first to realize that interdisciplinary collaborations are challenging and fraught with communication and cultural problems \cite{mnPb}, yet it seems that many organizations building ML-enabled systems pay little attention to fostering better interdisciplinary collaboration.

Organizations differ widely in their structures and practices, and some organizations have found strategies that work for them (see recommendation sections). Yet, we find that most organizations do not deliberately plan their structures and practices and have little insight into available choices and their tradeoffs. We hope that this work can (1) encourage more deliberation about organization and process at key collaboration points, and (2) serve as a starting point for cataloging and promoting best practices.

Beyond the specific challenges discussed throughout this paper, we see four broad themes that benefit from more attention both in engineering practice and in research:

\textbf{\faGroup\ Communication:} Many issues are rooted in miscommunication between participants with different backgrounds. To facilitate interdisciplinary collaboration, education is key, including ML literacy for software engineers and managers (and even customers) but also training data scientists to understand software engineering concerns. The idea of T-shaped professionals \cite{ULpd} (deep expertise in one area, broad knowledge of others) can provide guidance for hiring and training.

\newpage
\textbf{\faFileText\ Documentation:} Clearly documenting expectations between teams is important. Traditional interface documentation familiar to software engineers may be a starting point, but practices for documenting model requirements (Sec. \ref{h.6tv7b87udh5z}), data expectations (Sec. \ref{h.ng5qs5o69kjs}), and assured model qualities (Sec. \ref{h.ytgc9mej2ee2}) are not well established. Recent suggestions like model cards \cite{uiNH}, and FactSheets \cite{factsheet} are a good starting point for encouraging better, more standardized documentation of ML components. Given the interdisciplinary nature at these collaboration points, such documentation must be understood by all involved -- theories of \emph{boundary objects}~\cite{UVaE} may help to develop better interface description mechanisms.

\textbf{\faGears\ Engineering:} With attention focused on ML innovations, many organizations seem to underestimate the engineering effort required to turn a model into a product to be operated and maintained reliably. Arguably adopting machine learning increases software complexity \cite{eIMb,aGd5,ctC1} and makes engineering practices such as data quality checks, deployment automation, and testing in production even more important. Project managers should ensure that the ML and the non-ML parts of the project have sufficient engineering capabilities and foster product and operations thinking from the start.

\textbf{\faCalendar\ Process:} Finally, machine learning with its more science-like process challenges traditional software process life cycles. It seems clear that product requirements cannot be established without involving data scientists for model prototyping, and often it may be advisable to adopt a model-first trajectory to reduce risk. But while a focus on the product and overall process may cause delays, neglecting it entirely invites the kind of problems reported by our participants. Whether it may look more like the spiral model or agile \cite{sb6P}, more research into integrated process life cycles for ML-enabled systems (covering software engineering and data science) is needed.

\myparagraph{Acknowledgements.}
K\"astner's and Nahar's work was supported in part by NSF grants NSF award 1813598 and 2131477. Zhou's work was supported in part by Natural Sciences and Engineering Research Council of Canada (NSERC), RGPIN2021-03538. Lewis' work was funded and supported by the Department  of  Defense  under  Contract  No.  FA8702-15-D-0002 with Carnegie Mellon University (CMU) for the operation of the Software Engineering Institute (SEI), a federally funded research and development center. We would thank all our interview participants (K M Jawadur Rahman, Miguel Jette, and anonymous others) and the people who helped us connect with them.

\balance

\clearpage
\newpage
\title{Supplementary Material}
\label{supplement}
\maketitlewithoutauthor
\section*{Supplement A: Interview Participants}
\begin{table}[ht]
    \caption{Distribution of Company Location}
\begin{tabular}{@{}ll@{}}
\textbf{Company Location} & \multicolumn{1}{l}{\textbf{Count}} \\\hline
North America             & 11                                 \\\hline
South America             & 1                                  \\\hline
Europe                    & 5                                  \\\hline
Asia                      & 10                                 \\\hline
Africa                    & 1                                  \\\hline
                          & = 28                              
\end{tabular}
\end{table}
\begin{figure}[ht]
    \centering
\includegraphics[scale=0.5]{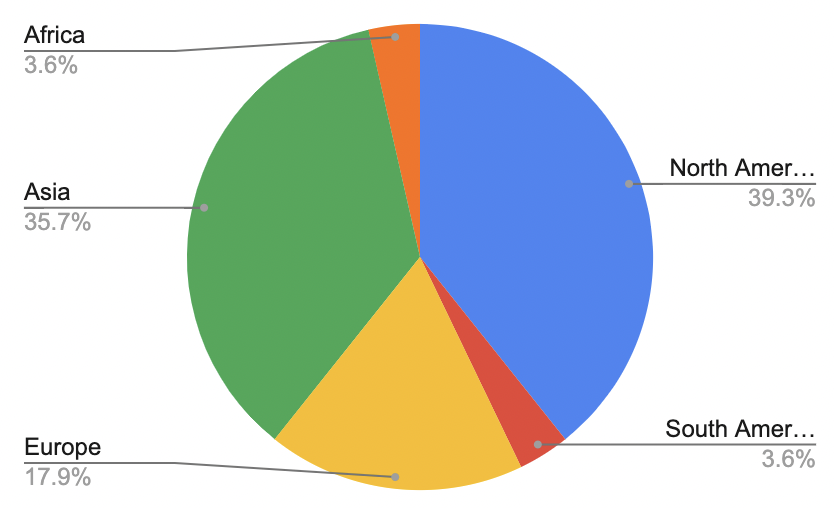}
    \caption{Distribution of Company Location}
    \label{fig:location}
\end{figure}

\FloatBarrier
\begin{table}[ht]
    \caption{Distribution of Participant Role}
\begin{tabular}{@{}ll@{}}
\textbf{Current Position} & \multicolumn{1}{l}{\textbf{Count}} \\\hline
ML-Related                & 23                                 \\\hline
SE-Related                & 9                                  \\\hline
Management                & 5                                  \\\hline
Operations                & 2                                  \\\hline
Domain Expert             & 2                                  \\\hline
Other                     & 4                                  \\\hline
                          & = 45                                
\end{tabular}
\end{table}

\begin{figure}[hb]
    \centering
\includegraphics[scale=0.47]{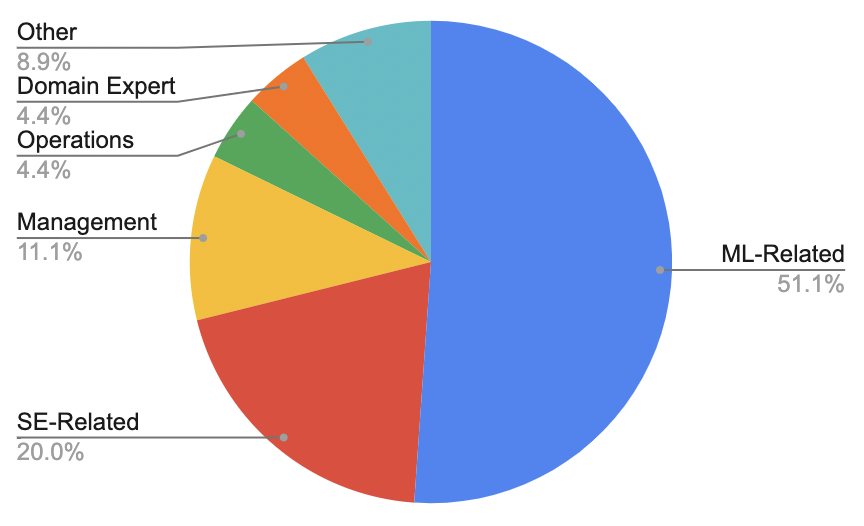}
    \caption{Distribution of Participant Role}
    \label{fig:role}
\end{figure}

\begin{table}[hb]
    \caption{Distribution of Company Type}
\begin{tabular}{@{}ll@{}}
\textbf{Company Type} & \multicolumn{1}{l}{\textbf{Count}} \\\hline
Big Tech              & 6                                  \\\hline
Non IT                & 4                                  \\\hline
Mid-size Tech         & 11                                 \\\hline
Startup               & 5                                  \\\hline
Consulting            & 2                                  \\\hline
                      & = 28                                
\end{tabular}
\end{table}

\begin{figure}[hb]
    \centering
\includegraphics[scale=0.47]{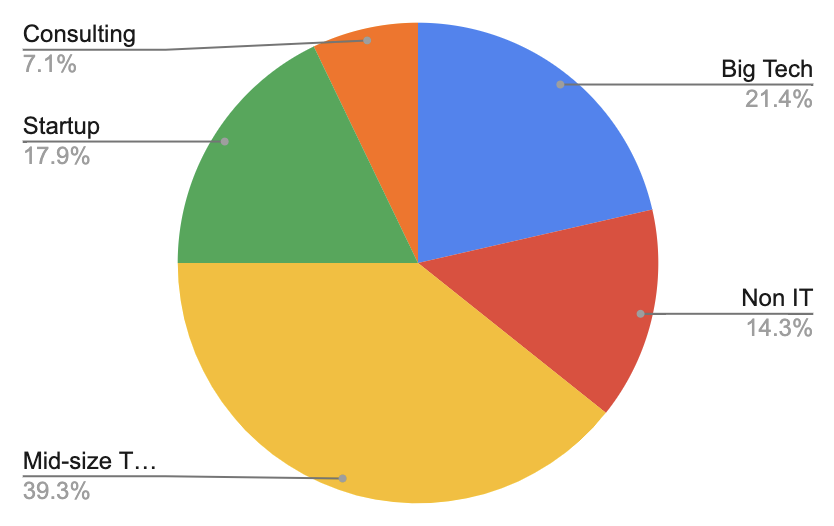}
    \caption{Distribution of Company Type}
    \label{fig:type}
\end{figure}

\clearpage
\newpage

\begin{table*}[ht]
\caption{Details of Interviewed Participants, Companies and Products}
\begin{tabular}{@{}p{0.1\textwidth}p{0.1\textwidth}p{0.1\textwidth}p{0.2\textwidth}p{0.4\textwidth}@{}}
\textbf{Company ID} &
  \textbf{Company Type} &
  \textbf{Participant ID} &
  \textbf{Participant Role} &
  \textbf{Product Domain} \\\hline
\multirow{2}{*}{g-1} &
  \multirow{2}{*}{Big Tech} &
  1a &
  SE &
  \multirow{2}{*}{NLP (entity recognition)} \\
 &
   &
  1b &
  Research Scientist (DS) &
   \\\hline
g-2 &
  Startup &
  2a &
  Co-founder (SE) &
  Nutrition app with personalized recommendation \\\hline
\multirow{3}{*}{g-3} &
  \multirow{3}{*}{Consulting} &
  3a &
  DS &
  \multirow{3}{*}{NLP model to identify pathogens} \\
 &
   &
  3b &
  SE and Operations &
   \\
 &
   &
  3c &
  SE &
   \\\hline
g-4 &
  Consulting &
  4a &
  SE &
  Video processing in edge devices \\\hline
\multirow{2}{*}{g-5} &
  \multirow{2}{*}{Big Tech} &
  5a &
  Research Scientist (SE) &
  \multirow{2}{*}{Scam detection} \\
 &
   &
  5b &
  Operations &
   \\\hline
\multirow{2}{*}{g-6}  & \multirow{2}{*}{Startup}       & 6a  & Co-founder (SE)             & \multirow{2}{*}{AI solution for communication feedback}               \\
 &
   &
  6b &
  AI scientiest &
   \\\hline
\multirow{2}{*}{g-7} &
  \multirow{2}{*}{Startup} &
  7a &
  DS researcher &
  \multirow{2}{*}{Circuit board analysis} \\
 &
   &
  7b &
  DS researcher &
   \\\hline
g-8 &
  Startup &
  8a &
  CEO &
  AI app for image analysis \\\hline
g-9 &
  Non IT &
  9a &
  Product manager &
  Autonomous car \\\hline
g-10 &
  Non IT &
  10a &
  DS &
  Warehouse forecasting \\\hline
g-11 &
  Mid-size Tech &
  11a &
  DS &
  Food recommendation \\\hline
g-12 &
  Non IT &
  12a &
  DS &
  Autonomous car \\\hline
g-13 &
  Startup &
  13a &
  DS &
  Price optimization for e-commerce \\\hline
\multirow{2}{*}{g-14} &
  \multirow{2}{*}{Mid-size Tech} &
  14a &
  DS &
  \multirow{2}{*}{Transcription service} \\
 &
   &
  14b &
  SE &
   \\\hline
\multirow{3}{*}{g-15} &
  \multirow{3}{*}{Mid-size Tech} &
  15a &
  Team lead (SE) &
  \multirow{3}{*}{OCR (Govt project)} \\
 &
   &
  15b &
  DS &
   \\
 &
   &
  15c &
  Consultant &
   \\\hline
\multirow{2}{*}{g-16} & \multirow{2}{*}{Mid-size Tech} & 16a & SE                          & \multirow{2}{*}{Speech detection (call center audio), chatbot}        \\
 &
   &
  16b &
  ML engineer &
   \\\hline
\multirow{2}{*}{g-17} &
  \multirow{2}{*}{Non IT} &
  17a &
  Data analyst &
  \multirow{2}{*}{Crop analysis (Govt project)} \\
 &
   &
  17b &
  Domain expert &
   \\\hline
\multirow{2}{*}{g-18} & \multirow{2}{*}{Mid-size Tech} & 18a & SE (work in both SE and ML) & \multirow{2}{*}{Chat-bot (multiple client including banks)}            \\
 &
   &
  18b &
  DS &
   \\\hline
\multirow{2}{*}{g-19} & \multirow{2}{*}{Mid-size Tech} & 19a & Team lead (ML)              & \multirow{2}{*}{Pharmaceutical's product}                               \\
 &
   &
  19b &
  ML and AI engineer &
   \\\hline
\multirow{2}{*}{g-20} & \multirow{2}{*}{Mid-size Tech} & 20a & ML engineer                 & \multirow{2}{*}{Identify prescribed medicine from prescription image} \\
 &
   &
  20b &
  ML engineer &
   \\\hline
\multirow{2}{*}{g-21} & \multirow{2}{*}{Mid-size Tech} & 21a & CEO and Architect           & \multirow{2}{*}{Video processing app using existing AI engine}        \\
 &
   &
  21b &
  SE &
   \\\hline
\multirow{3}{*}{g-22} & \multirow{3}{*}{Mid-size Tech} & 22a & COO                         & \multirow{3}{*}{Audio processing for keyword detection}               \\
 &
   &
  22b &
  Analytics &
   \\
 &
   &
  22c &
  DS &
   \\\hline
g-23 &
  Mid-size Tech &
  23a &
  Lead Data Scientist &
  Dengue Early Warning System \\\hline
g-24 &
  Mid-size Tech &
  24a &
  SE + DS &
  Video processing app using existing AI engine \\\hline
g-25 &
  Big Tech &
  25a &
  Data Scientist &
  Text classification \\\hline
g-26 &
  Big Tech &
  26a &
  Senior Data Scientist &
  OS crash report analysis \\\hline
g-27 &
  Big Tech &
  27a &
  Senior Data Scientist &
  Draud detection \\\hline
g-28 &
  Big Tech &
  28a &
  Staff Software Engineer (managing cross ML team) &
  E-commerce product classification \\\hline
\end{tabular}
\end{table*}
\clearpage
\section*{Supplement B: Interview Guide}

\subsection*{Introductory Comments}
\begin{itemize}
\item[--] This study intended to understand the challenges of collaboration between different stakeholders in a machine learning production system, and the current industry best practices to deal with those challenges. In support of this study, a series of targeted interviews are being conducted to gather information from the stakeholders involved in different stages of building, deploying, and operating machine learning production systems.
\item[--] Please do not share any confidential information with us. If you prefer information not to be included in the recording, just let us know. 
\item[--] The information gathered from the interviews will be aggregated in the form of mismatches and consequences, and will have no connection to any organization.
\item[--] Before we begin, we would like to notify you that we intend to record the interview for transcription purposes. Only the research team will be able to access the recordings and transcripts. After the interview is transcribed and identifying information is removed from the transcript, the audio recording will be destroyed. 
\item[--] The audio recording will be done in a private space for both physical and over phone conversations.
\end{itemize}

\subsection*{Interview Questions}
\subsubsection*{[Intro]}
\begin{description}
\item[Project and Role] Can you tell me more about you; your role at your company, and your academic and professional background? Tell me about the last ml-based project you worked on. To what extent you were involved in the system, and what was/were your roles in it?
\item[ML Component] Can you tell me about the ml or non-ml parts of the project? How the ML component is used as part of the larger system? 
\item[Team] Can you describe a bit about your team? What are the roles involved? What are their tasks? And how do they communicate with each other, and for what purpose? Who is involved with which components?
\end{description}
\subsubsection*{[Topic-Specific Questions]}
\begin{description}
\item[Understanding system requirements and ML capabilities] Who were involved in the decision to use ML in this project (or whether to turn an ML prototype into a product)? Who do you talk to for the requirements of the system as well as the ML components in the project? -- (probe) ask about feedback loops
\item[Project planning and Process] 
\begin{description}
\item[Project planning] How do you plan or estimate for the project, specifically the ML components and dependencies between components? Who is involved in the planning? Are these documented? 
\item[Process] In what order are the things developed in the ML project? Do you follow a process model? 
\item[Dealing with change] Do you plan for change management? How does this planning / replanning happen? Who gets involved?
\end{description}
\item[Components] Can you tell me a bit about your ML pipeline? What are the interaction points to the non-ML components? Can you describe/draw the architecture of the system with ml and non-ml components?
\item[System decomposition] How did you decide to decompose the whole system into dependent/independent components this way? Was it obvious or did it change? Do you think of ML as one component? Who was involved in the decomposition process and deciding the module boundaries? 
\item[System Correctness] How do you plan to evaluate your system correctness? How do you set quality goals?
\begin{description}
\item[ML component testing] How do you plan to test the model? Do you conduct offline/online testing? What data is collected later (e.g., for telemetry)? Who made these decisions and do they evolve?
\item[System testing] In testing, how much do you focus on testing the model versus testing the entire system? How do you test the system? Who is involved in testing? Who makes decisions? 
\item[Breaking change] Did you face any application break recently? What is considered a breaking change? How do you deal with breaking changes? How do you detect them? -- (probe) Do you consider that model might evolve during model development and are likely to influence other parts of the system? 
\end{description}
\item[Data quality] How much do you or others in the project worry about data quality?
\begin{description}

\item[Data need] Think about data that you receive from other parts of the system (or from outside the system) or of data you produce for other parts of the system: Are data quality and quantity requirements documented? Are there schema definitions or data quality checks or monitoring? Who "owns" the data? Who cleans the data? Who is responsible for checks or documentation? What happens when data format or quality change? How much influence do you have on data quality and quantity?
\item[Data understanding] Is the data schema and its semantics documented somewhere? Who does the documentation? If you don't understand the data, whom do you communicate with to understand it?
\item[Planning and monitoring for drift] What happens if the data changes? Who is responsible for notifying changes? How do you get to know about the change? How do you deal with schema evolution? 
\end{description}
\item[Special qualities] Do you consider requirements like explainability or fairness, and legal requirements like privacy? Are the requirements documented at the system or model level? Who is responsible? Do you plan for fairness/ robustness testing? 
\item[Versioning] Do you maintain model versioning? What about data versioning? Do you think about provenance tracking? Who made decisions? Who is affected?
\item[Reusability, Reproducibility, Maintainability] To what extent do you develop documentation for the component? Do you follow any coding standards or conventions? Do you consider designing the module to reuse? Do you care whether components and  results are reproducible? 
\item[Operations] How do you deploy updates to models and non-ML components? How often? Who? Do you consider continuous experimentation? Who is responsible for the operations? 
\item[Composition/Integration] How do you transfer the model from prototype to production? Who is responsible for integrating the ML model to the system? Do multiple roles collaborate for this or is it done by one specific role?  What is the difficulty level for that, according to you? 
\end{description}
\subsubsection*{[Last Thoughts]}
\begin{description}
\item[ML vs non-ML] Can you share your experience with the ML project in comparison with other types of projects that do not include a machine learning component (if you have worked on any)?
\item[Challenges/ Benefits] Can you think about any challenges you faced during the project development or afterward? Why do you think the challenge arose, and in which step? -- (probe) [Interdisciplinary Collaboration] Is working with team members of different backgrounds a challenge in this project? 
\end{description}
\section*{Supplement C: Codebook}

\textbf{\emph{1. Understanding System Requirements}}

\emph{Description}: This is the collaboration point where the broad system requirements are collected/defined. The overall system goal and system requirements may differ from those of individual ML components and focus on the overall system behavior and its interactions with the environment. While a system can grow around keeping an ML module as a central point, another system can consist of less important ML module(s) working along with other traditional features. Therefore, system requirements may be collected upfront or late and incrementally after the initial model development. For the first category of the system, the requirements are generated after the ML component is defined. Thus, the ML design decisions influence how the entire system is shaped, and take control on defining the requirements of the overall system. For the second category of system, the requirements gathering is executed similarly as a traditional software. One or few ML features are incorporated into the overall system, that are influenced and shaped by the broad system-level requirements constraining the data scientists during modeling. Based on these categories, we expect that stakeholders think differently about the business needs of the system. In this collaboration point, the stakeholders need to define the system scope, system-level requirements and metrics to set quality expectations for the overall system. This includes considering how the system interacts with the environment and how this influences safety, security, fairness, and feedback-loop issues at the system level.  Also, this is the collaboration point where the special requirements like explainability, fairness, privacy, safety, security, Human-AI interaction planning, etc. need to be defined properly for the overall system as well as realize their effect on the ML part. In short, the different stakeholders of the system need to be involved in the requirements collection and realization process.

\emph{Agree on}: System-level requirements, reasoning about feedback loops, non-functional requirements of the system

\emph{Output}: Documentation on requirements including functional-nonfunctional requirements, environment interaction analysis reasoning about feedback loops.

\emph{Involved}: Primarily - Requirements Engineer, Domain Expert, Customer and Manager. Needs consultation with - Software Engineer, Data Scientist, Tester, Operator.  

\emph{Lifecycle}: Traditionally, this would happen in early requirements stages of a waterfall-like process or iteratively in other process models. In ML projects this may happen when shifting from initial modeling to building a production system.

\emph{Why challenging}: While developing ML components, the data scientists often forget about the overall system and focus on the model part only. Thus, inconsistencies may arise if the model does not correspond to the system goals. Also, it is often reported that stakeholders find it difficult to define the scope of the project that includes ML components. ML components often lead to false expectations that makes it difficult to set reasonable targets. Additionally, it’s difficult to quantify the quality targets or the qualitative non-functional requirements. Identifying the feedback loops is also not a trivial task. Moreover, as both the traditional and ML parts are involved and the team consists of people with heterogeneous languages and priorities, these people need to interact with each other to come to a common position.

\emph{Example}: Google photos app has ML modules for photo tagging. However, the app has a lot of other features, and the ML module needs to interact and be consistent with those. For example - should the app show the photo tags to the users or will the users be able to change a tag if they think its incorrect. Thus, the different stakeholders of the Google photos app need to be involved in the requirements collection and realization process.
 
\textbf{\emph{2. Project Planning, Process and Interdisciplinary Collaboration}}

\emph{Description}: This is the collaboration point that deals with the overall project planning. It defines the process model to be followed. While a project might start with collecting requirements first, another can start with letting some data scientists play with data for a year. So the process needs to be determined at this point. Additionally, general planning like time estimation, risk mitigation plan etc. also need to be conducted and synchronized among the stakeholders at this collaboration point. As this collaboration point is where the overall system planning and process is discussed, the whole team involved in the project should be involved here, at least implicitly. That is why, the communication challenges between the interdisciplinary team members is also combined with this point.

\emph{Agree on}: Project plan and process models to be followed. Communication agendas between different teams/roles.

\emph{Output}: Informal project planning to more formal planning documents, possible process documentation, adoption of process practices, forming of interdisciplinary teams, adoption of team work practices

\emph{Involved}: Primarily - Project Manager. Needs consultation with - Software Engineer, Data Scientist, Domain Experts, Tester, Operator, etc.

\emph{Lifecycle}: Generally, in a waterfall-like process, the tasks can be actively handled in both System Requirements and Planning/ High-level Design (outside ML pipeline) stages. However, in practice, it generally spreads out and continues as ongoing tasks. 

\emph{Why challenging}: In general, time estimation is hard for the ML applications due to their exploratory nature. This also creates a lot of collaboration challenges in planning due to the differences between SE and ML components. 
Example: Google photos app has ML modules and other traditional functionalities. The ML modules have structures and time requirements different from the other traditional parts. Thus, before building the app, the project planning needs to incorporate planning for the ML parts. For example - the photo tagging component might need a different timeline than the traditional components. A risk analysis might be required for the component, and similar to the iterative model, the risk-associated component can be attempted to build first.
 
\textbf{\emph{3. System Decomposition, Local Checking and System Evaluation}}

\emph{Description}: This collaboration point deals with decomposing the overall system into ML and non-ML SE components. This defines the module boundaries and negotiates the requirements or interface contracts for each of the components. While each component will be evaluated internally against its interfaces, also later composition and integration-/system-level quality assurance is planned, ensuring that the composed system meets the system specifications. During decomposition and local-system evaluation many qualities need to be considered, each resulting possibly with corresponding obligations at the module interfaces. Generally, we consider four kinds of components:
\begin{itemize}
\item non-ML component: These are the traditional SE components
\item Pipeline component: This component represents the process of producing the model, roughly how data is transferred to the model and the deployment of it.
\item Inference component: This is the component that is concerned about the model prediction, and thus, relates to using the model. 
\item Monitoring component: This represents the component for monitoring the system after the deployment of the system.
\end{itemize}

\emph{Agree on}: Module boundaries and quality requirements for the components after the decomposition. Also, evaluation mechanisms for the components and the overall system after the composition. 
 
\textbf{\emph{3.1. Functional Correctness / Target Domain / Fit (Accuracy)}}

\emph{Description}: Functional "correctness" is used as a broad term here, that measures if the system produces outputs that match a given problem. The system has behavioral expectations that need to be broken down into individual components, tested locally, and then be composed and evaluated again at the system level. 

\emph{Agree on}: Accuracy expectations for ML components, functional behavior expectations of all individual components, test strategy for components, test strategies for integration

\emph{Component-level Evaluation}: Model Accuracy. Both Offline/ Online testing can be needed. Telemetry collection needs collaboration. Functional correctness of non-ML components.

\emph{Involved}: Primarily - Data Scientist. For online testing/telemetry collection, needs consultation with - Domain Expert, Software Engineer, Operator, Tester.

\emph{System-level Evaluation}: Integration, system and acceptance testing.

\emph{Involved}: Primarily - Tester. Needs consultation with - Software Engineer, Data Scientist, Operator, Manager.
 
\textbf{\emph{3.2. Fairness, Privacy and Accountability}}

\emph{Description}: Inclusion of ML in a system often creates expectations of some special system-level quality requirements that includes - fairness, privacy, explainability, provenance tracking, etc. Again, to achieve the quality at the system level, the role of individual components needs to be negotiated, defined, and assured; the integration of components needs to be evaluated.

\emph{Agree on}: Protected data points, level of fairness expected, level of explainability/provenance tracking expected

\emph{Component-level Evaluation}: Model Fairness Testing (increasing data quantity might be needed),  Constraints on Protected Values in Data, Use of Explainable Algorithms, Versioning Code, Model and Data.

\emph{Involved}: Primarily - Data Scientist. Needs consultation with - Domain Expert, Legal Department, Manager, Tester.

\emph{System-level Evaluation}: System-level Fairness Testing, Testing whether provenance tracking can be performed, Testing whether a system output can be explained or whether the results are interpretable.

\emph{Involved}: Primarily - Tester. Needs consultation with - Software Engineer, Data Scientist, Domain Expert, Legal Department, Operator, Manager.
 
\textbf{\emph{3.3. Data Quality}}

\emph{Description}: Data is one of the most important parts of an ML application. This includes tracking data needs both from the aspect of quality and quantity, defining an appropriate data schema, monitoring data drifts and taking necessary steps to incorporate change or evolution, documentation of data, etc. 
Agree on: Quality and quantity of data to be collected, data schema definition, data understanding and reporting, data monitoring and managing drifts

\emph{Component-level Evaluation}: Improve/Degrade of Accuracy can be one indicator of data needs. Privacy is a concern here as data can contain protected attributes. Data cleaning and storing in schema are steps to check on data integrity in mode level.

\emph{Involved}: Primarily - Data Scientist. Needs consultation with - Legal Department [for privacy], Domain Expert [for schema definition and drift], Manager [for more data need].

\emph{System-level Evaluation}: Degrade of model accuracy in the integrated system can be one indicator of data drift.

\emph{Involved}: Primarily - Data Scientist. Needs consultation with - Software Engineer, Tester, Operator [for collecting telemetry], Domain Expert [for drift], Manager [for more data need].
 
\textbf{\emph{3.4. Reusability, Reproducibility, Maintainability, Infrastructure Quality}}

\emph{Description}: ML components have different coding languages and different conventions. However, for a complete software, the parts might follow similar coding conventions so that the other team members can understand the code easily and integrate the parts together without much effort. Also, the documentation of the implementation needs to be maintained to a certain extent so that the system can be easily updated later on. This will help to increase the system maintainability. This also leads to easy reusability and reproducibility of the system.

\emph{Agree on}: Coding Convention and Level of Documentation, Expectations of Reusability, Reproducibility and Maintainability

\emph{Component-level Evaluation}: Consistent Coding and Documentation of the Component\hphantom{~~~~}(\emph{Involved}: Primarily - Software Engineer and Data Scientist. Needs consultation with - Manager)

\emph{System-level Evaluation}: Consistent Coding and Documentation throughout the System.

\emph{Involved}: Primarily - Software Engineer and Data Scientist. Needs consultation with - Manager.
 
\textbf{\emph{3.5. Updatability, Operations}}

\emph{Description}: As ML systems need to be monitored for data drifts and online testing with telemetry collection, deployment is not the last stage of development. Operations is an important part of such projects leading to the popularity of MLOps. Along with the data monitoring, change updates are also required for both ML and traditional parts of the software. Continuous experimentation is another aspect that requires updating the application continuously. This relates to system versioning and provenance tracking as well.

\emph{Agree on}: Deployment and Monitoring Mechanisms, Frequency of Updates, Continuous Experimentation Requirements

\emph{Component-level Evaluation}: Stable Model Deployment and Monitoring Mechanisms for Data Drifts. Experimentation with Different Versions of the Model.\hphantom{~~~~}(\emph{Involved}: Primarily - Software Engineer and Data Scientist. Needs consultation with - Operator, Manager and Tester)

\emph{System-level Evaluation}: Stable System Deployment and Monitoring Mechanisms. Monitoring for change requests and continuous bug fixes generate continuous patch or version updates.

\emph{Involved}: Primarily - Operator. Needs consultation with - Software Engineer, Tester and Manager.
 
\textbf{\emph{3.6. Usability}}

\emph{Description}: Like the other traditional systems, the ML systems also need user interaction in different forms. Whatever the form is, user satisfaction/usability is an important aspect of any system. Especially, ML systems need a collection of telemetry that demands extra attention to the UI design.

\emph{Agree on}: User Experience Expectations

\emph{Involved}: Primarily - Software Engineer and UX Expert. Needs consultation with - Data Scientist, Operator, Manager, and Tester


\section*{Supplement D: List of Papers}

\subsection*{Initial Set of Papers (15)}

\begingroup
\renewcommand{\section}[2]{}%
\begin{itemize}
\item Chattopadhyay, S., Prasad, I., Henley, A. Z., Sarma, A., \& Barik, T. (2020). What's Wrong with Computational Notebooks? Pain Points, Needs, and Design Opportunities. In \emph{Proceedings of the 2020 CHI Conference on Human Factors in Computing Systems} (pp. 1–12). Association for Computing Machinery.
\item O'Leary, K., \& Uchida, M. (2020). \emph{Common problems with creating machine learning pipelines from existing code}. 
\item Li, P. L., Ko, A. J., \& Begel, A. (2017). Cross-Disciplinary Perspectives on Collaborations with Software Engineers. In \emph{2017 IEEE/ACM 10th International Workshop on Cooperative and Human Aspects of Software Engineering (CHASE)}.
\item Kim, M., Zimmermann, T., DeLine, R., \& Begel, A. (2018). Data Scientists in Software Teams: State of the Art and Challenges. In \emph{IEEE Transactions on Software Engineering} (Vol. 44, Issue 11, pp. 1024–1038).
\item Arpteg, A., Brinne, B., Crnkovic-Friis, L., \& Bosch, J. (2018). Software Engineering Challenges of Deep Learning. \emph{2018 44th Euromicro Conference on Software Engineering and Advanced Applications (SEAA)}, 50–59.
\item Breck, E., Cai, S., Nielsen, E., Salib, M., \& Sculley, D. (2017). The ML test score: A rubric for ML production readiness and technical debt reduction. \emph{2017 IEEE International Conference on Big Data (Big Data)}, 1123–1132.
\item Kery, M. B., Radensky, M., Arya, M., John, B. E., \& Myers, B. A. (2018). The story in the notebook. \emph{Proceedings of the 2018 CHI Conference on Human Factors in Computing Systems - CHI '18}. the 2018 CHI Conference, Montreal QC, Canada.
\item Zhang, A. X., Muller, M., \& Wang, D. (2020). How do data science workers collaborate? Roles, workflows, and tools. \emph{Proceedings of the ACM on Human-Computer Interaction}, \emph{4}(CSCW1), 1–23.
\item Pimentel, J. F., Murta, L., Braganholo, V., \& Freire, J. (2019, May). A large-scale study about quality and reproducibility of jupyter notebooks. \emph{2019 IEEE/ACM 16th International Conference on Mining Software Repositories (MSR)}. 2019 IEEE/ACM 16th International Conference on Mining Software Repositories (MSR), Montreal, QC, Canada.
\item Head, A., Hohman, F., Barik, T., Drucker, S. M., \& DeLine, R. (2019). Managing messes in computational notebooks. \emph{Proceedings of the 2019 CHI Conference on Human Factors in Computing Systems - CHI '19}. the 2019 CHI Conference, Glasgow, Scotland Uk. 
\item Studer, S., Bui, T. B., Drescher, C., Hanuschkin, A., Winkler, L., Peters, S., \& Mueller, K.-R. (2020). Towards CRISP-ML(Q): A Machine Learning Process Model with Quality Assurance Methodology. In \emph{arXiv [cs.LG]}. arXiv. 
\item Sculley, D., Holt, G., Golovin, D., Davydov, E., Phillips, T., Ebner, D., Chaudhary, V., Young, M., Crespo, J.-F., \& Dennison, D. (2015). Hidden Technical Debt in Machine Learning Systems. In C. Cortes, N. D. Lawrence, D. D. Lee, M. Sugiyama, \& R. Garnett (Eds.), \emph{Advances in Neural Information Processing Systems 28} (pp. 2503–2511). Curran Associates, Inc.
\item Amershi, S., Begel, A., Bird, C., DeLine, R., Gall, H., Kamar, E., Nagappan, N., Nushi, B., \& Zimmermann, T. (2019). Software Engineering for Machine Learning: A Case Study. In \emph{2019 IEEE/ACM 41st International Conference on Software Engineering: Software Engineering in Practice (ICSE-SEIP)}.
\item Polyzotis, N., Roy, S., Whang, S. E., \& Zinkevich, M. (2017). Data Management Challenges in Production Machine Learning. \emph{Proceedings of the 2017 ACM International Conference on Management of Data}, 1723–1726.
\item Vogelsang, A., \& Borg, M. (2019). Requirements Engineering for Machine Learning: Perspectives from Data Scientists. \emph{2019 IEEE 27th International Requirements Engineering Conference Workshops (REW)}, 245–251.

\end{itemize}

\subsection*{Complete Set of Papers (61)}

\begin{itemize}
\item Chattopadhyay, S., Prasad, I., Henley, A. Z., Sarma, A., \& Barik, T. (2020). What's Wrong with Computational Notebooks? Pain Points, Needs, and Design Opportunities. In \emph{Proceedings of the 2020 CHI Conference on Human Factors in Computing Systems} (pp. 1–12). Association for Computing Machinery.
\item O'Leary, K., \& Uchida, M. (2020). \emph{Common problems with creating machine learning pipelines from existing code}. 
\item Li, P. L., Ko, A. J., \& Begel, A. (2017). Cross-Disciplinary Perspectives on Collaborations with Software Engineers. In \emph{2017 IEEE/ACM 10th International Workshop on Cooperative and Human Aspects of Software Engineering (CHASE)}.
\item Kim, M., Zimmermann, T., DeLine, R., \& Begel, A. (2018). Data Scientists in Software Teams: State of the Art and Challenges. In \emph{IEEE Transactions on Software Engineering} (Vol. 44, Issue 11, pp. 1024–1038).
\item Arpteg, A., Brinne, B., Crnkovic-Friis, L., \& Bosch, J. (2018). Software Engineering Challenges of Deep Learning. \emph{2018 44th Euromicro Conference on Software Engineering and Advanced Applications (SEAA)}, 50–59.
\item Breck, E., Cai, S., Nielsen, E., Salib, M., \& Sculley, D. (2017). The ML test score: A rubric for ML production readiness and technical debt reduction. \emph{2017 IEEE International Conference on Big Data (Big Data)}, 1123–1132.
\item Kery, M. B., Radensky, M., Arya, M., John, B. E., \& Myers, B. A. (2018). The story in the notebook. \emph{Proceedings of the 2018 CHI Conference on Human Factors in Computing Systems - CHI '18}. the 2018 CHI Conference, Montreal QC, Canada.
\item Zhang, A. X., Muller, M., \& Wang, D. (2020). How do data science workers collaborate? Roles, workflows, and tools. \emph{Proceedings of the ACM on Human-Computer Interaction}, \emph{4}(CSCW1), 1–23.
\item Pimentel, J. F., Murta, L., Braganholo, V., \& Freire, J. (2019, May). A large-scale study about quality and reproducibility of jupyter notebooks. \emph{2019 IEEE/ACM 16th International Conference on Mining Software Repositories (MSR)}. 2019 IEEE/ACM 16th International Conference on Mining Software Repositories (MSR), Montreal, QC, Canada.
\item Head, A., Hohman, F., Barik, T., Drucker, S. M., \& DeLine, R. (2019). Managing messes in computational notebooks. \emph{Proceedings of the 2019 CHI Conference on Human Factors in Computing Systems - CHI '19}. the 2019 CHI Conference, Glasgow, Scotland Uk. 
\item Studer, S., Bui, T. B., Drescher, C., Hanuschkin, A., Winkler, L., Peters, S., \& Mueller, K.-R. (2020). Towards CRISP-ML(Q): A Machine Learning Process Model with Quality Assurance Methodology. In \emph{arXiv [cs.LG]}. arXiv. 
\item Sculley, D., Holt, G., Golovin, D., Davydov, E., Phillips, T., Ebner, D., Chaudhary, V., Young, M., Crespo, J.-F., \& Dennison, D. (2015). Hidden Technical Debt in Machine Learning Systems. In C. Cortes, N. D. Lawrence, D. D. Lee, M. Sugiyama, \& R. Garnett (Eds.), \emph{Advances in Neural Information Processing Systems 28} (pp. 2503–2511). Curran Associates, Inc.
\item Amershi, S., Begel, A., Bird, C., DeLine, R., Gall, H., Kamar, E., Nagappan, N., Nushi, B., \& Zimmermann, T. (2019). Software Engineering for Machine Learning: A Case Study. In \emph{2019 IEEE/ACM 41st International Conference on Software Engineering: Software Engineering in Practice (ICSE-SEIP)}.
\item Polyzotis, N., Roy, S., Whang, S. E., \& Zinkevich, M. (2017). Data Management Challenges in Production Machine Learning. \emph{Proceedings of the 2017 ACM International Conference on Management of Data}, 1723–1726.
\item Vogelsang, A., \& Borg, M. (2019). Requirements Engineering for Machine Learning: Perspectives from Data Scientists. \emph{2019 IEEE 27th International Requirements Engineering Conference Workshops (REW)}, 245–251.
\item Holstein, K., Wortman Vaughan, J., Daumé, H., Dudik, M., \& Wallach, H. (2019). Improving Fairness in Machine Learning Systems: What Do Industry Practitioners Need? \emph{Proceedings of the 2019 CHI Conference on Human Factors in Computing Systems}, 1–16.
\item Rahimi, M., Guo, J. L. C., Kokaly, S., \& Chechik, M. (2019). Toward Requirements Specification for Machine-Learned Components. \emph{2019 IEEE 27th International Requirements Engineering Conference Workshops (REW)}, 241–244.
\item Nushi, B., Kamar, E., Horvitz, E., \& Kossmann, D. (2017). On human intellect and machine failures: troubleshooting integrative machine learning systems. \emph{Proceedings of the Thirty-First AAAI Conference on Artificial Intelligence}, 1017–1025.
\item Borg, M., Englund, C., Wnuk, K., Duran, B., Levandowski, C., Gao, S., Tan, Y., Kaijser, H., Lönn, H., \& Törnqvist, J. (2019). Safely Entering the Deep: A Review of Verification and Validation for Machine Learning and a Challenge Elicitation in the Automotive Industry. In \emph{Journal of Automotive Software Engineering} (Vol. 1, Issue 1, p. 1). 
\item Kandel, S., Paepcke, A., Hellerstein, J. M., \& Heer, J. (2012). Enterprise data analysis and visualization: An interview study. \emph{IEEE Transactions on Visualization and Computer Graphics}, \emph{18}(12), 2917–2926.
\item Madaio, M. A., Stark, L., Wortman Vaughan, J., \& Wallach, H. (2020). Co-Designing Checklists to Understand Organizational Challenges and Opportunities around Fairness in AI. \emph{Proceedings of the 2020 CHI Conference on Human Factors in Computing Systems}, 1–14.
\item Salay, R., Queiroz, R., \& Czarnecki, K. (2017). An Analysis of ISO 26262: Using Machine Learning Safely in Automotive Software. In \emph{arXiv [cs.AI]}. arXiv.
\item Ashmore, R., Calinescu, R., \& Paterson, C. (2019). Assuring the Machine Learning Lifecycle: Desiderata, Methods, and Challenges. In \emph{arXiv [cs.LG]}. arXiv. 
\item Sendak, M. P., Ratliff, W., Sarro, D., Alderton, E., Futoma, J., Gao, M., Nichols, M., Revoir, M., Yashar, F., Miller, C., Kester, K., Sandhu, S., Corey, K., Brajer, N., Tan, C., Lin, A., Brown, T., Engelbosch, S., Anstrom, K., … O'Brien, C. (2020). Real-World Integration of a Sepsis Deep Learning Technology Into Routine Clinical Care: Implementation Study. \emph{JMIR Medical Informatics}, \emph{8}(7), e15182.
\item Sculley, D., Otey, M. E., Pohl, M., Spitznagel, B., Hainsworth, J., \& Zhou, Y. (2011). Detecting adversarial advertisements in the wild. \emph{Proceedings of the 17th ACM SIGKDD International Conference on Knowledge Discovery and Data Mining}, 274–282.
\item Bernardi, L., Mavridis, T., \& Estevez, P. (2019). 150 successful machine learning models. \emph{Proceedings of the 25th ACM SIGKDD International Conference on Knowledge Discovery \& Data Mining - KDD '19}. the 25th ACM SIGKDD International Conference, Anchorage, AK, USA.
\item Lwakatare, L. E., Raj, A., Bosch, J., Olsson, H. H., \& Crnkovic, I. (2019). A taxonomy of software engineering challenges for machine learning systems: An empirical investigation. \emph{International Conference on Agile Software Development}, 227–243.
\item Yang, Q., Suh, J., Chen, N.-C., \& Ramos, G. (2018). Grounding Interactive Machine Learning Tool Design in How Non-Experts Actually Build Models. \emph{Proceedings of the 2018 Designing Interactive Systems Conference}, 573–584.
\item Martinez-Plumed, F., Contreras-Ochando, L., Ferri, C., Hernandez Orallo, J., Kull, M., Lachiche, N., Ramirez Quintana, M. J., \& Flach, P. A. (2020). CRISP-DM twenty years later: From data mining processes to data science trajectories. \emph{IEEE Transactions on Knowledge and Data Engineering}, 1–1.
\item Ishikawa, F. and Yoshioka, N., 2019, May. How do engineers perceive difficulties in engineering of machine-learning systems?-questionnaire survey. In 2019 IEEE/ACM Joint \emph{7th International Workshop on Conducting Empirical Studies in Industry (CESI) and 6th International Workshop on Software Engineering Research and Industrial Practice (SER\&IP) (pp. 2-9)}. IEEE.
\item Ozkaya, I. (2020). What Is Really Different in Engineering AI-Enabled Systems? \emph{IEEE Software}, \emph{37}(4), 3–6.
\item Wiens, J., Saria, S., Sendak, M., Ghassemi, M., Liu, V. X., Doshi-Velez, F., Jung, K., Heller, K., Kale, D., Saeed, M., Ossorio, P. N., Thadaney-Israni, S., \& Goldenberg, A. (2019). Do no harm: a roadmap for responsible machine learning for health care. \emph{Nature Medicine}, \emph{25}(9), 1337–1340.
\item Wagstaff, K. (2012). Machine Learning that Matters. In \emph{arXiv [cs.LG]}. arXiv.
\item Hynes, N., Sculley, D. and Terry, M., 2017. The data linter: Lightweight, automated sanity checking for ml data sets. \emph{In NIPS MLSys Workshop}.
\item Strubell, E., Ganesh, A. and McCallum, A., 2019. Energy and policy considerations for deep learning in NLP. arXiv preprint \emph{arXiv:1906.02243}.
\item Hukkelberg, I., \& Rolland, K. (2020). \emph{EXPLORING MACHINE LEARNING IN A LARGE GOVERNMENTAL ORGANIZATION: AN INFORMATION INFRASTRUCTURE PERSPECTIVE}. 
\item Hill, C., Bellamy, R., Erickson, T. and Burnett, M., 2016, September. Trials and tribulations of developers of intelligent systems: A field study. \emph{In 2016 IEEE Symposium on Visual Languages and Human-Centric Computing (VL/HCC) (pp. 162-170)}. 
\item Baylor, D., Breck, E., Cheng, H.-T., Fiedel, N., Foo, C. Y., Haque, Z., Haykal, S., Ispir, M., Jain, V., Koc, L., Koo, C. Y., Lew, L., Mewald, C., Modi, A. N., Polyzotis, N., Ramesh, S., Roy, S., Whang, S. E., Wicke, M., … Zinkevich, M. (n.d.). \emph{TFX: A TensorFlow-Based Production-Scale Machine Learning Platform}.
\item Ré, C., Niu, F., Gudipati, P. and Srisuwananukorn, C., 2019. Overton: A data system for monitoring and improving machine-learned products. arXiv preprint \emph{arXiv:1909.05372}.
\item Bhatt, U., Xiang, A., Sharma, S., Weller, A., Taly, A., Jia, Y., Ghosh, J., Puri, R., Moura, J. M. F., \& Eckersley, P. (2020). Explainable machine learning in deployment. \emph{Proceedings of the 2020 Conference on Fairness, Accountability, and Transparency}, 648–657.
\item Hazelwood, K., Bird, S., Brooks, D., Chintala, S., Diril, U., Dzhulgakov, D., Fawzy, M., Jia, B., Jia, Y., Kalro, A. and Law, J., 2018, February. Applied machine learning at facebook: A datacenter infrastructure perspective. \emph{In 2018 IEEE International Symposium on High Performance Computer Architecture (HPCA) (pp. 620-629)}. IEEE.
\item Amershi, S., Chickering, M., Drucker, S. M., Lee, B., Simard, P., \& Suh, J. (2015). ModelTracker: Redesigning Performance Analysis Tools for Machine Learning. \emph{Proceedings of the 33rd Annual ACM Conference on Human Factors in Computing Systems}, 337–346.
\item Peng, Z., Yang, J., Chen, T.-H. (peter), \& Ma, L. (2020, November 8). A first look at the integration of machine learning models in complex autonomous driving systems: a case study on Apollo. \emph{Proceedings of the 28th ACM Joint Meeting on European Software Engineering Conference and Symposium on the Foundations of Software Engineering}. ESEC/FSE '20: 28th ACM Joint European Software Engineering Conference and Symposium on the Foundations of Software Engineering, Virtual Event USA.
\item Polyzotis, N., Roy, S., Whang, S. E., \& Zinkevich, M. (2018). Data Lifecycle Challenges in Production Machine Learning: A Survey. \emph{SIGMOD Rec.}, \emph{47}(2), 17–28.
\item Humbatova, N., Jahangirova, G., Bavota, G., Riccio, V., Stocco, A., \& Tonella, P. (2020, June 27). Taxonomy of real faults in deep learning systems. \emph{Proceedings of the ACM/IEEE 42nd International Conference on Software Engineering}. ICSE '20: 42nd International Conference on Software Engineering, Seoul South Korea. 
\item Polyzotis, N., Zinkevich, M., Roy, S., Breck, E., \& Whang, S. (2019). Data validation for machine learning. \emph{Proceedings of Machine Learning and Systems}, \emph{1}, 334–347.
\item Wan, Z., Xia, X., Lo, D., \& Murphy, G. C. (2019). How does Machine Learning Change Software Development Practices? \emph{IEEE Transactions on Software Engineering}, 1–1.
\item Lwakatare, L. E., Raj, A., Crnkovic, I., Bosch, J., \& Olsson, H. H. (2020). Large-scale machine learning systems in real-world industrial settings: A review of challenges and solutions. \emph{Information and Software Technology}, \emph{127}(106368), 106368.
\item Siebert, J., Joeckel, L., Heidrich, J., Nakamichi, K., Ohashi, K., Namba, I., Yamamoto, R., \& Aoyama, M. (2020). Towards Guidelines for Assessing Qualities of Machine Learning Systems. In \emph{Communications in Computer and Information Science} (pp. 17–31). 
\item Shneiderman, B. (2020). Bridging the gap between ethics and practice. \emph{ACM Transactions on Interactive Intelligent Systems}, \emph{10}(4), 1–31.
\item Seymoens, T., Ongenae, F., \& Jacobs, A. (2018). A methodology to involve domain experts and machine learning techniques in the design of human-centered algorithms. \emph{Working Conference on …}. 
\item Zinkevich, M. (2017). Rules of machine learning: Best practices for ML engineering. \emph{URL: Https://developers. Google. Com/machine-Learning/guides/rules-of-Ml}. 
\item Park, S., Wang, A., Kawas, B., Vera Liao, Q., Piorkowski, D., \& Danilevsky, M. (2021). Facilitating Knowledge Sharing from Domain Experts to Data Scientists for Building NLP Models. In \emph{arXiv [cs.HC]}. arXiv. 
\item d. S. Nascimento, E., Ahmed, I., Oliveira, E., Palheta, M. P., Steinmacher, I., \& Conte, T. (2019). Understanding Development Process of Machine Learning Systems: Challenges and Solutions. \emph{2019 ACM/IEEE International Symposium on Empirical Software Engineering and Measurement (ESEM)}, 1–6.
\item Cai, C. J., Winter, S., Steiner, D., Wilcox, L., \& Terry, M. (2019). ``hello AI'': Uncovering the onboarding needs of medical practitioners for human-AI collaborative decision-making. \emph{Proceedings of the ACM on Human-Computer Interaction}, \emph{3}(CSCW), 1–24.
\item Amershi, S., Weld, D., Vorvoreanu, M., Fourney, A., Nushi, B., Collisson, P., Suh, J., Iqbal, S., Bennett, P. N., Inkpen, K., Teevan, J., Kikin-Gil, R., \& Horvitz, E. (2019). Guidelines for Human-AI Interaction. \emph{Proceedings of the 2019 CHI Conference on Human Factors in Computing Systems}, 1–13.
\item Piorkowski, D., Park, S., Wang, A. Y., Wang, D., Muller, M., \& Portnoy, F. (2021). How AI Developers Overcome Communication Challenges in a Multidisciplinary Team: A Case Study. In \emph{arXiv [cs.CY]}. arXiv. 
\item Haakman, M., Cruz, L., Huijgens, H., \& van Deursen, A. (2020). AI Lifecycle Models Need To Be Revised. An Exploratory Study in Fintech. In \emph{arXiv [cs.SE]}. arXiv. 
\item McGraw, G., Figueroa, H., Shepardson, V. and Bonett, R., 2020. An architectural risk analysis of machine learning systems: Toward more secure machine learning. Berryville Institute of Machine Learning, Clarke County, VA. Accessed on: Mar, 23.
\item Hopkins, A., \& Booth, S. (2021, July 21). Machine learning practices outside big tech: How resource constraints challenge responsible development. \emph{Proceedings of the 2021 AAAI/ACM Conference on AI, Ethics, and Society}. AIES '21: AAAI/ACM Conference on AI, Ethics, and Society, Virtual Event USA. 
\item Rakova, B., Yang, J., Cramer, H., \& Chowdhury, R. (2021). Where Responsible AI meets Reality: Practitioner Perspectives on Enablers for Shifting Organizational Practices. \emph{Proc. ACM Hum.-Comput. Interact.}, \emph{5}(CSCW1), 1–23.
\end{itemize}

\clearpage
\newpage


































































































































































































































\begin{thebibliography}{100}
\bibitem{aho} Aho, T., Sievi-Korte, O., Kilamo, T., Yaman, S. and Mikkonen, T., 2020. Demystifying data science projects: A look on the people and process of data science today. \emph{In Proc. Int'l Conf. Product-Focused Software Process Improvement}, 153-167. 
\bibitem{UVaE} Akkerman, S.F. and Bakker, A. 2011. Boundary Crossing and Boundary Objects. \emph{Review of educational research}. 81, 2, 132–169.
\bibitem{VVrZ} Akkiraju, R., Sinha, V., Xu, A., Mahmud, J., Gundecha, P., Liu, Z., Liu, X. and Schumacher, J. 2020. Characterizing Machine Learning Processes: A Maturity Framework. \emph{Business Process Management}, 17–31.
\bibitem{I17M} Ameisen, E. 2020. \emph{Building Machine Learning Powered Applications: Going from Idea to Product}. O'Reilly Media, Inc.
\bibitem{tPQ5} Amershi, S. et al. 2019. Software Engineering for Machine Learning: A Case Study. \emph{In Proc. of 41st Int'l Conf. on Software Engineering: Software Engineering in Practice (ICSE-SEIP)}, 291–300.
\bibitem{x4oR} Amershi, S., Chickering, M., Drucker, S.M., Lee, B., Simard, P. and Suh, J. 2015. ModelTracker: Redesigning Performance Analysis Tools for Machine Learning. \emph{In Proc. of 33rd Conf. on Human Factors in Computing Systems}, 337–346.
\bibitem{TIPJ} Amershi, S. et al. 2019. Guidelines for Human-AI Interaction. \emph{In Proc. of CHI Conf. on Human Factors in Computing Systems}, 1–13.
\bibitem{factsheet} Arnold, M. et al. 2019. FactSheets: Increasing trust in AI services through supplier's declarations of conformity. \emph{IBM Journal of Research and Development}, 63.
\bibitem{9g1T} Arpteg, A., Brinne, B., Crnkovic-Friis, L. and Bosch, J. 2018. Software Engineering Challenges of Deep Learning. \emph{In Proc. Euromicro Conf. Software Engineering and Advanced Applications (SEAA)}, 50–59.
\bibitem{Pd6Q} Ashmore, R., Calinescu, R. and Paterson, C. 2021. Assuring the Machine Learning Lifecycle: Desiderata, Methods, and Challenges. \emph{ACM Computing Surveys (CSUR)}, 54 (5): 1-39.
\bibitem{FvQJ} Bass, L., Clements, P. and Kazman, R. 1998. \emph{Software Architecture in Practice}. Addison-Wesley Longman Publishing Co., Inc.
\bibitem{Ler4} Bass, M., Herbsleb, J.D. and Lescher, C. 2009. A Coordination Risk Analysis Method for Multi-site Projects: Experience Report. \emph{In Proc. Int'l Conf. Global Software Engineering}, 31–40.
\bibitem{6aXp} Baylor, D., Breck, E., Cheng, H.T., Fiedel, N., Foo, C.Y. et al. 2017. TFX: A TensorFlow-Based Production-Scale Machine Learning Platform. \emph{In Proc. Int'l Conf. Knowledge Discovery and Data Mining}, 1387-1395.
\bibitem{2kVg} Bernardi, L., Mavridis, T. and Estevez, P. 2019. 150 successful machine learning models. \emph{In Proc. Int'l Conf. Knowledge Discovery \& Data Mining}, 1743-1751.
\bibitem{Yuwe} Bhatt, U., Xiang, A., Sharma, S., Weller, A., Taly, A., Jia, Y., Ghosh, J., Puri, R., Moura, J.M.F. and Eckersley, P. 2020. Explainable machine learning in deployment. \emph{In Proc. of Conf. on Fairness, Accountability, and Transparency}, 648–657.
\bibitem{1i9g} Bogart, C., Kästner, C., Herbsleb, J. and Thung, F. 2021. When and how to make breaking changes: Policies and practices in 18 open source software ecosystems. \emph{ACM Transactions on Software Engineering and Methodology}. 30, 4, 1–56.
\bibitem{We4r} Borg, M. et al. 2019. Safely Entering the Deep: A Review of Verification and Validation for Machine Learning and a Challenge Elicitation in the Automotive Industry. \emph{Journal of Automotive Software Engineering}. 1, 1, 1–9.
\bibitem{6ckN} Bosch, J., Olsson, H.H. and Crnkovic, I. 2021. Engineering AI Systems: A Research Agenda. \emph{Artificial Intelligence Paradigms for Smart Cyber-Physical Systems}. IGI Global. 1–19.
\bibitem{zu7r} Boujut, J.-F. and Blanco, E. 2003. Intermediary Objects as a Means to Foster Co-operation in Engineering Design. \emph{Computer Supported Cooperative Work (CSCW)}. 12, 2, 205–219.
\bibitem{i8ow} Braiek, H.B. and Khomh, F. 2020. On testing machine learning programs. \emph{Journal of Systems and Software}. 164, 110542.
\bibitem{mnPb} Brandstädter, S. and Sonntag, K. 2016. Interdisciplinary Collaboration. \emph{Advances in Ergonomic Design of Systems, Products and Processes}, 395–409.
\bibitem{sb6P} Braude, Eric J and Bernstein, Michael E. 2011. \emph{Software Engineering: Modern Approaches 2nd Edition}. Wiley. \emph{ISBN-13: 978-0471692089}.
\bibitem{Lmxi} Breck, E., Cai, S., Nielsen, E., Salib, M. and Sculley, D. 2017. The ML test score: A rubric for ML production readiness and technical debt reduction. \emph{In Proc. of Int'l Conf. on Big Data (Big Data)}, 1123–1132.
\bibitem{N1oU} Brown, G.F.C. 1995. \emph{Factors that facilitate or inhibit interdisciplinary collaboration within a professional bureaucracy}. University of Arkansas.
\bibitem{uoFv} Cai, C.J., Winter, S., Steiner, D., Wilcox, L. and Terry, M. 2019. ``hello AI'': Uncovering the onboarding needs of medical practitioners for human-AI collaborative decision-making. \emph{In Proc. Human-Computer Interaction}. 3, CSCW, 1–24.
\bibitem{2BW8} Cataldo, M. et al. 2006. Identification of Coordination Requirements: Implications for the Design of Collaboration and Awareness Tools. \emph{In Proc. Conf. Computer Supported Cooperative Work (CSCW)}, 353–362.
\bibitem{7sej} Chattopadhyay, S., Prasad, I., Henley, A.Z., Sarma, A. and Barik, T. 2020. What's Wrong with Computational Notebooks? Pain Points, Needs, and Design Opportunities. \emph{In Proc. Conf. Human Factors in Computing Systems (CHI)}, 1–12.
\bibitem{bQ59} Cheng, D., Cao, C., Xu, C. and Ma, X. 2018. Manifesting Bugs in Machine Learning Code: An Explorative Study with Mutation Testing. \emph{In Proc. Int'l Conf. Software Quality, Reliability and Security (QRS)}, 313–324.
\bibitem{5pvj} Chen, Z., Cao, Y., Liu, Y., Wang, H., Xie, T. and Liu, X. 2020. Understanding Challenges in Deploying Deep Learning Based Software: An Empirical Study. \emph{arXiv 2005.00760}.
\bibitem{mFZV} Conway, M.E. 1968. How Do Committees Invent? \emph{Datamation}. 14, 4, 28–31.
\bibitem{4JFg} Cossette, B.E. and Walker, R.J. 2012. Seeking the Ground Truth: A Retroactive Study on the Evolution and Migration of Software Libraries. \emph{In Proc. Int'l Symposium Foundations of Software Engineering (FSE)}, 1–11.
\bibitem{JztF} Curtis, B., Krasner, H. and Iscoe, N. 1988. A field study of the software design process for large systems. \emph{Communications of the ACM}. 31, 11, 1268–1287.
\bibitem{MuoQ} Dabbish, L., Stuart, C., Tsay, J. and Herbsleb, J. 2012. Social Coding in GitHub: Transparency and Collaboration in an Open Software Repository. \emph{In Proc. Conf. Computer Supported Cooperative Work (CSCW)}, 1277–1286.
\bibitem{5guS} Haakman, M., Cruz, L., Huijgens, H. and van Deursen, A. 2021. AI Lifecycle Models Need To Be Revised. An exploratory study in Fintech. \emph{Empirical Software Engineering}. 26, 5, 1–29.
\bibitem{o88o} Haakman, M., Cruz, L., Huijgens, H. and van Deursen, A. 2020. AI Lifecycle Models Need To Be Revised. An Exploratory Study in Fintech. \emph{arXiv 2010.02716}.
\bibitem{c7gw} Harsh, S. 2011. Purposeful Sampling in Qualitative Research Synthesis. \emph{Qualitative Research Journal}. 11, 2, 63–75.
\bibitem{4sUH} Head, A., Hohman, F., Barik, T., Drucker, S.M. and DeLine, R. 2019. Managing messes in computational notebooks. \emph{In Proc. Conf. Human Factors in Computing Systems (CHI)}, 1-12.
\bibitem{FlkT} Herbsleb, J.D. and Grinter, R.E. 1999. Splitting the Organization and Integrating the Code: Conway's Law Revisited. \emph{In Proc.  Int'l Conf. Software Engineering (ICSE)}, 85–95.
\bibitem{sWPF} Holstein, K. et al.. 2019. Improving Fairness in Machine Learning Systems: What Do Industry Practitioners Need? \emph{In Proc.  Conf. Human Factors in Computing (CHI) Systems}, 1–16.
\bibitem{gC0G} Hopkins, A. and Booth, S. 2021. Machine learning practices outside big tech: How resource constraints challenge responsible development. \emph{In Proc. Conf. on AI, Ethics, and Society}, 134-145.
\bibitem{K8oj} Hukkelberg, I. and Rolland, K. 2020. Exploring Machine Learning in a Large Governmental Organization: An Information Infrastructure Perspective. \emph{European Conf. on Information Systems}, 92.
\bibitem{hpbm} Hulten, G. 2019. \emph{Building Intelligent Systems: A Guide to Machine Learning Engineering}. Apress.
\bibitem{efZw} Humbatova, N. et al. 2020. Taxonomy of real faults in deep learning systems. \emph{In Proc. Int'l Conf. on Software Engineering (ICSE)}, 1110-1121.
\bibitem{FdO8} Ishikawa, F. and Yoshioka, N. 2019. How do engineers perceive difficulties in engineering of machine-learning systems? - Questionnaire survey. \emph{In Proc. Int'l Workshop on Conducting Empirical Studies in Industry (CESI) and Software Engineering Research and Industrial Practice (SER\&IP)}, 2-9.
\bibitem{gXfW} Islam, M.J., Nguyen, H.A., Pan, R. and Rajan, H. 2019. What Do Developers Ask About ML Libraries? A Large-scale Study Using Stack Overflow. \emph{arXiv 1906.11940}.
\bibitem{hzdi} Kandel, S., Paepcke, A., Hellerstein, J.M. and Heer, J. 2012. Enterprise data analysis and visualization: An interview study. \emph{IEEE Transactions on Visualization and Computer Graphics}. 18, 12, 2917–2926.
\bibitem{kj2C} Kang, D., Raghavan, D., Bailis, P. and Zaharia, M. 2020. Model Assertions for Monitoring and Improving ML Models. \emph{In Proc. of Machine Learning and Systems, 2, 481-496}.
\bibitem{eIMb} Kästner, C. and Kang, E. 2020. Teaching Software Engineering for Al-Enabled Systems. \emph{In Proc. Int'l Conf. Software Engineering: Software Engineering Education and Training (ICSE-SEET)}, 45–48.
\bibitem{JnkY} Kim, M., Zimmermann, T., DeLine, R. and Begel, A. 2018. Data Scientists in Software Teams: State of the Art and Challenges. \emph{IEEE Transactions on Software Engineering}. 44, 11, 1024–1038.
\bibitem{kuwajima} Kuwajima, H., Yasuoka, H. and Nakae, T. 2020. Engineering problems in machine learning systems. \emph{Machine Learning}, 109, no 5, 1103-1126.
\bibitem{Kol9} Lakshmanan, V., Robinson, S. and Munn, M. 2020. \emph{Machine Learning Design Patterns}. O'Reilly Media, Inc.
\bibitem{9nnb} Lewis, G.A., Bellomo, S. and Ozkaya, I. 2021. Characterizing and Detecting Mismatch in Machine-Learning-Enabled Systems. \emph{In Proc. Workshop on AI Engineering-Software Engineering for AI (WAIN), 133-140}.
\bibitem{lewis2021} Lewis, G. A., Ozkaya, I. and Xu X. 2021. Software Architecture Challenges for ML Systems. \emph{In Proc. Int'l Conf. on Software Maintenance and Evolution}, 634-638.
\bibitem{7wRo} Li, P.L., Ko, A.J. and Begel, A. 2017. Cross-Disciplinary Perspectives on Collaborations with Software Engineers. \emph{In Proc. Int'l Workshop on Cooperative and Human Aspects of Software Engineering (CHASE)}, 2–8.
\bibitem{q6tY} Lwakatare, L.E., Raj, A., Bosch, J., Olsson, H.H. and Crnkovic, I. 2019. A taxonomy of software engineering challenges for machine learning systems: An empirical investigation. \emph{In Proc. Int'l Conf. Agile Software Development}, 227–243.
\bibitem{Mfyd} Lwakatare, L.E., Raj, A., Crnkovic, I., Bosch, J. and Olsson, H.H. 2020. Large-scale machine learning systems in real-world industrial settings: A review of challenges and solutions. \emph{Information and software technology}. 127, 106368.
\bibitem{Emzk} Madaio, M.A. et al. 2020. Co-Designing Checklists to Understand Organizational Challenges and Opportunities around Fairness in AI. \emph{In Proc. Conf. Human Factors in Computing Systems (CHI)}, 1–14.
\bibitem{jt5V} Mahanti, R. 2019. \emph{Data Quality: Dimensions, Measurement, Strategy, Management, and Governance}. Quality Press.
\bibitem{5hJ9} Mäkinen, S., Skogström, H., Laaksonen, E. and Mikkonen, T. 2021. Who Needs MLOps: What Data Scientists Seek to Accomplish and How Can MLOps Help? \emph{In Proc. Workshop on AI Engineering-Software Engineering for AI (WAIN)}, 109-112.
\bibitem{NhgE} Martínez-Fernández, S., Bogner, J., Franch, X., Oriol, M., Siebert, J., Trendowicz, A., Vollmer, A.M. and Wagner, S. 2021. Software Engineering for AI-Based Systems: A Survey. \emph{arXiv 2105.01984}.
\bibitem{B5qD} Martinez-Plumed, F., Contreras-Ochando, L., Ferri, C., Hernandez Orallo, J., Kull, M., Lachiche, N., Ramirez Quintana, M.J. and Flach, P.A. 2021. CRISP-DM twenty years later: From data mining processes to data science trajectories. \emph{IEEE Transactions on Knowledge and Data Engineering}. 33, 8, 3048–3061.
\bibitem{06Iv} Meyer, B. 1997. \emph{Object-Oriented Software Construction}. Prentice-Hall.
\bibitem{hATC} Mistrík, I., Grundy, J., van der Hoek, A. and Whitehead, J. 2010. \emph{Collaborative Software Engineering}. Springer.
\bibitem{uiNH} Mitchell, M., Wu, S., Zaldivar, A., Barnes, P., Vasserman, L., Hutchinson, B., Spitzer, E., Raji, I.D. and Gebru, T. 2019. Model Cards for Model Reporting. \emph{In Proc. Conf. Fairness, Accountability, and Transparency}, 220–229.
\bibitem{muiruri} Muiruri, D., Lwakatare, L. E., K Nurminen, J. and Mikkonen, T. 2021. Practices and Infrastructures for ML Systems--An Interview Study. \emph{TechRxiv 16939192.v1}.
\bibitem{ZfhU} O'Leary, K. and Uchida, M. 2020. Common problems with creating machine learning pipelines from existing code. \emph{In Proc. Conf. Machine Learning and Systems (MLSys)}.
\bibitem{AzHU} Ovaska, P., Rossi, M. and Marttiin, P. 2003. Architecture as a coordination tool in multi-site software development. \emph{Software Process Improvement and Practice}. 8, 4, 233–247.
\bibitem{ctC1} Ozkaya, I. 2020. What Is Really Different in Engineering AI-Enabled Systems? \emph{IEEE Software}. 37, 4, 3–6.
\bibitem{GYwY} Panetta, K. 2020. Gartner Identifies the Top Strategic Technology Trends for 2021. \emph{URL: \url{https://www.gartner.com/smarterwithgartner/gartner-top-strategic-technology-trends-for-2021}}.
\bibitem{42j1} Park, S., Wang, A., Kawas, B., Vera Liao, Q., Piorkowski, D. and Danilevsky, M. 2021. Facilitating Knowledge Sharing from Domain Experts to Data Scientists for Building NLP Models. \emph{In Proc. 26th Int'l Conf. on Intelligent User Interfaces}, 585-596.
\bibitem{VII1} Parnas, D.L. 1972. On the Criteria to be used in Decomposing Systems into Modules. \emph{Communications of the ACM}. 15, 12, 1053–1058.
\bibitem{passi} Passi, S., and Phoebe S. 2020. Making Data Science Systems Work. \emph{Big Data \& Society} 7 (2): 1-13.
\bibitem{sIUl} Patel, K., Fogarty, J., Landay, J.A. and Harrison, B. 2008. Investigating statistical machine learning as a tool for software development. \emph{In Proc. Conf. Human Factors in Computing Systems (CHI)}, 667–676.
\bibitem{MAdB} Pimentel, J.F., Murta, L., Braganholo, V. and Freire, J. 2019. A large-scale study about quality and reproducibility of Jupyter notebooks. \emph{In Proc. 16th Int'l Conf. on Mining Software Repositories (MSR)}, 507-517.
\bibitem{evYm} Piorkowski, D. et al. 2021. How AI Developers Overcome Communication Challenges in a Multidisciplinary Team: A Case Study. \emph{In Proc. ACM on Human-Computer Interaction}, 5, (CSCW1), 1-25.
\bibitem{CY3n} Polyzotis, N., Roy, S., Whang, S.E. and Zinkevich, M. 2018. Data Lifecycle Challenges in Production Machine Learning: A Survey. \emph{SIGMOD Rec.} 47, 2, 17–28.
\bibitem{JN0G} Polyzotis, N., Roy, S., Whang, S.E. and Zinkevich, M. 2017. Data Management Challenges in Production Machine Learning. \emph{In Proc. Int'l Conf. on Management of Data}, 1723–1726.
\bibitem{7QTl} Polyzotis, N., Zinkevich, M., Roy, S., Breck, E. and Whang, S. 2019. Data validation for machine learning. \emph{In Proc. Machine Learning and Systems}, 334–347.
\bibitem{ExZ5} Rahimi, M., Guo, J.L.C., Kokaly, S. and Chechik, M. 2019. Toward Requirements Specification for Machine-Learned Components. \emph{In Proc. Int'l Requirements Engineering Conf. Workshops (REW)}, 241–244.
\bibitem{p7yN} Rakova, B., Yang, J., Cramer, H. and Chowdhury, R. 2021. Where Responsible AI meets Reality: Practitioner Perspectives on Enablers for Shifting Organizational Practices. \emph{Proc. ACM Hum.-Comput. Interact.} 5, CSCW1, 1–23.
\bibitem{gWib} Ré, C., Niu, F., Gudipati, P. and Srisuwananukorn, C. 2019. Overton: A data system for monitoring and improving machine-learned products. \emph{arXiv 1909.05372}.
\bibitem{vhGB} Salay, R., Queiroz, R. and Czarnecki, K. 2017. An Analysis of ISO 26262: Using Machine Learning Safely in Automotive Software. \emph{arXiv 1709.02435}.
\bibitem{WVIq} Sambasivan, N. et al. 2021. ``Everyone wants to do the model work, not the data work'': Data Cascades in High-Stakes AI. \emph{In Proc. Conf. on Human Factors in Computing Systems (CHI)}.  1–15.
\bibitem{03Lt} Sarma, A., Redmiles, D.F. and van der Hoek, A. 2012. Palantir: Early Detection of Development Conflicts Arising from Parallel Code Changes. \emph{IEEE Transactions on Software Engineering}. 38, 4, 889–908.
\bibitem{ubVS} Schelter, S et al. 2018. Automating Large-scale Data Quality Verification. \emph{Proc. VLDB Endowment Int'l Conf. Very Large Data Bases}. 11, 12, 1781–1794.
\bibitem{aGd5} Sculley, D. et al. 2015. Hidden Technical Debt in Machine Learning Systems. \emph{Advances in Neural Information Processing Systems 28}. 2503–2511.
\bibitem{sAxw} Sculley, D., Otey, M.E., Pohl, M., Spitznagel, B., Hainsworth, J. and Zhou, Y. 2011. Detecting adversarial advertisements in the wild. \emph{In Proc. Int'l Conf. Knowledge Discovery and Data Mining}, 274–282.
\bibitem{9t4Z} Sendak, M.P. et al. 2020. Real-World Integration of a Sepsis Deep Learning Technology Into Routine Clinical Care: Implementation Study. \emph{JMIR medical informatics}. 8, 7, e15182.
\bibitem{UjZF} Serban, A., van der Blom, K., Hoos, H. and Visser, J. 2020. Adoption and Effects of Software Engineering Best Practices in Machine Learning. \emph{In Proc. Int'l Symposium on Empirical Software Engineering and Measurement}, 1–12.
\bibitem{yL64} Seymoens, T., Ongenae, F. and Jacobs, A. 2018. A methodology to involve domain experts and machine learning techniques in the design of human-centered algorithms. \emph{In Proc. IFIP Working Conf. Human Work Interaction Design}, 200-214.
\bibitem{7dgT} Shneiderman, B. 2020. Bridging the gap between ethics and practice. \emph{ACM Transactions on Interactive Intelligent Systems}. 10, 4, 1–31.
\bibitem{Dzd7} Siebert, J., Joeckel, L., Heidrich, J., Nakamichi, K., Ohashi, K., Namba, I., Yamamoto, R. and Aoyama, M. 2020. Towards Guidelines for Assessing Qualities of Machine Learning Systems. \emph{In Proc. Int'l Conf. on the Quality of Information and Communications Technology}, 17–31.
\bibitem{XRxi} Singh, G., Gehr, T., Püschel, M. and Vechev, M. 2019. An abstract domain for certifying neural networks. \emph{Proc. ACM Program. Lang.} 3, POPL, 1–30.
\bibitem{ozxl} Smith, D., Alshaikh, A., Bojan, R., Kak, A. and Manesh, M.M.G. 2014. Overcoming barriers to collaboration in an open source ecosystem. \emph{Technology Innovation Management Review}. 4, 1.
\bibitem{XeyT} d. S. Nascimento, E. et al. 2019. Understanding Development Process of Machine Learning Systems: Challenges and Solutions. \emph{In Proc. Int'l Symposium on Empirical Software Engineering and Measurement (ESEM)}, 1–6.
\bibitem{GqTH} de Souza, C.R.B. and Redmiles, D.F. 2008. An Empirical Study of Software Developers' Management of Dependencies and Changes. \emph{In Proc. Int'l Conf. Software Engineering (ICSE)}, 241–250.
\bibitem{Hc0J} Strauss, A. and Corbin, J. 1994. Grounded theory methodology: An overview. \emph{Handbook of qualitative research}. N.K. Denzin, ed. 273–285.
\bibitem{ia1C} Strauss, A. and Corbin, J.M. \emph{Basics of Qualitative Research: Grounded Theory Procedures and Techniques}. SAGE Publications.
\bibitem{kckt} Studer, S. et al. 2021. Towards CRISP-ML(Q): A Machine Learning Process Model with Quality Assurance Methodology. \emph{Machine Learning and Knowledge Extraction}, 3(2), 392-413.
\bibitem{8A28} Tramèr, F. et al. 2017. FairTest: Discovering Unwarranted Associations in Data-Driven Applications. \emph{In Proc. European Symposium on Security and Privacy (EuroS P)}, 401–416.
\bibitem{ULpd} Tranquillo, J. 2017. The T-Shaped Engineer. \emph{Journal of Engineering Education Transformations}. 30, 4, 12–24.
\bibitem{aFTn} Vogelsang, A. and Borg, M. 2019. Requirements Engineering for Machine Learning: Perspectives from Data Scientists. \emph{In Proc. Int'l Requirements Engineering Conf. Workshops (REW)}, 245–251.
\bibitem{8e8c} Wagstaff, K. 2012. Machine Learning that Matters. \emph{arXiv 1206.4656}.
\bibitem{U2s5} Wang, A.Y., Mittal, A., Brooks, C. and Oney, S. 2019. How Data Scientists Use Computational Notebooks for Real-Time Collaboration. \emph{Proc. Human-Computer Interaction}. 3, CSCW, 39.
\bibitem{bU0N} Wan, Z., Xia, X., Lo, D. and Murphy, G.C. 2019. How does Machine Learning Change Software Development Practices? \emph{IEEE Transactions on Software Engineering}, 47(9), 1857-1871.
\bibitem{WTXV} Waterman, M., Noble, J. and Allan, G. 2015. How Much Up-Front? A Grounded theory of Agile Architecture. \emph{In Proc. Int'l Conf. Software Engineering}, 347–357.
\bibitem{sr9P} Staff, V. B. 2019. Why do 87\% of data science projects never make it into production? \emph{URL: \url{https://venturebeat.com/2019/07/19/why-do-87-of-data-science-projects-never-make-it-into-production/}}.
\bibitem{ynnp} Wiens, J., et al. 2019. Do no harm: A roadmap for responsible machine learning for health care. \emph{Nature medicine}. 25, 9, 1337–1340.
\bibitem{WaYh} Xie, X., Ho, J.W.K., Murphy, C., Kaiser, G., Xu, B. and Chen, T.Y. 2011. Testing and Validating Machine Learning Classifiers by Metamorphic Testing. \emph{Journal of Systems and Software}. 84, 4, 544–558.
\bibitem{5Nhu} Yang, Q., Suh, J., Chen, N.-C. and Ramos, G. 2018. Grounding Interactive Machine Learning Tool Design in How Non-Experts Actually Build Models. \emph{In Proc. Conf. Designing Interactive Systems}, 573–584.
\bibitem{yang} Yang, Q. The role of design in creating machine-learning-enhanced user experience. \emph{In Proc. AAAI Spring Symposium Series}, 406-411.
\bibitem{bdD7} Yokoyama, H. 2019. Machine Learning System Architectural Pattern for Improving Operational Stability. \emph{In Proc. Int'l Conf. on Software Architecture Companion (ICSA-C)}, 267–274.
\bibitem{7LH3} Zhang, A.X., Muller, M. and Wang, D. 2020. How do data science workers collaborate? Roles, workflows, and tools. \emph{Proc. Human-Computer Interaction}. 4, CSCW1, 1–23.
\bibitem{pevg} Zhou, S., Vasilescu, B. and Kästner, C. 2020. How Has Forking Changed in the Last 20 Years? A Study of Hard Forks on GitHub. \emph{In Proc. Int'l Conf. Software Engineering (ICSE)}, 445–456.
\bibitem{Jmep} Zinkevich, M. 2017. Rules of machine learning: Best practices for ML engineering. \emph{URL: \url{https://developers.google.com/machine-learning/guides/rules-of-ml}}.
\end{thebibliography}
\end{document}